\documentclass[sigconf]{acmart}

\usepackage{tabularx}
\usepackage{multicol}
\usepackage{multirow}
\usepackage{booktabs}
\usepackage{algorithm}
\usepackage[noend]{algpseudocode}
\usepackage{makecell}
\usepackage{pgfplots}
\usepackage{paralist}
\usepackage{amsmath}
\usepackage{listings}
\usepackage{tikz}
\usepackage{etoolbox}
\usepackage{enumitem}
\usepackage{subcaption}
\usetikzlibrary{arrows}
\usetikzlibrary{shapes.geometric}
\usetikzlibrary{shapes.multipart}

\algnewcommand{\IfThenElse}[3]{
  \algorithmicif\ #1\ \algorithmicthen\ #2\ \algorithmicelse\ #3}
\DeclareMathOperator*{\argmin}{argmin}

\author{Abhinav Jangda}
\email{aabhinav@cs.umass.edu}
\affiliation{%
  \institution{University of Massachusetts Amherst}
  \country{United States}
}

\author{Arjun Guha}
\email{a.guha@northeastern.edu}
\affiliation{%
  \institution{Northeastern University}
  \country{United States}
}

\begin{document}
\fancyhead{}

\copyrightyear{2020}
\acmYear{2020}
\setcopyright{acmcopyright}\acmConference[PACT '20]{Proceedings of the 2020 International Conference on Parallel Architectures and Compilation Techniques}{October 3--7, 2020}{Virtual Event, GA, USA}
\acmBooktitle{Proceedings of the 2020 International Conference on Parallel Architectures and Compilation Techniques (PACT '20), October 3--7, 2020, Virtual Event, GA, USA}
\acmPrice{15.00}
\acmDOI{10.1145/3410463.3414649}
\acmISBN{978-1-4503-8075-1/20/10}

\ccsdesc[500]{Software and its engineering~Compilers}
  
\keywords{Polyhedral Optimizations; Graphics Processing Units; Image Processing Pipelines}

\title{Model-Based Warp Overlapped Tiling for Image Processing Programs on GPUs}

\newcommand{\nsms}{\texttt{NSMs}}
\newcommand{\ShPerSM}{\texttt{ShMemPerSM}}
\newcommand{\RegPerSM}{\texttt{RegPerSM}}
\newcommand{\MaxRegPerThread}{\texttt{MaxRegPerTh}}
\newcommand{\MaxWarpPerSM}{\texttt{MaxWarpPerSM}}
\newcommand{\GLMemBW}{\texttt{GlMemBW}}
\newcommand{\warpSize}{\texttt{WarpSize}}
\newcommand{\MaxShPerTB}{\texttt{MaxShMemPerTb}}
\newcommand{\MaxThreadBlockPerSM}{\texttt{MaxTbPerSM}}
\newcommand{\MaxThreadsPerSM}{\texttt{MaxThPerSM}}
\newcommand{\NCudaCoresPerSM}{\texttt{CoresPerSM}}
\newcommand{\GLMemTxSize}{\texttt{GlMemTxSz}}

\newcounter{groupcount}
\pgfplotsset{
    draw group line/.style n args={5}{
        after end axis/.append code={
            \setcounter{groupcount}{0}
            \pgfplotstableforeachcolumnelement{#1}\of\datatable\as\cell{%
                \def\temp{#2}
                \ifx\temp\cell
                    \ifnum\thegroupcount=0
                        \stepcounter{groupcount}
                        \pgfplotstablegetelem{\pgfplotstablerow}{[index]0}\of\datatable
                        \coordinate [yshift=#4] (startgroup) at (axis cs:\pgfplotsretval,0);
                    \else
                        \pgfplotstablegetelem{\pgfplotstablerow}{[index]0}\of\datatable
                        \coordinate [yshift=#4] (endgroup) at (axis cs:\pgfplotsretval,0);
                    \fi
                \else
                    \ifnum\thegroupcount=1
                        \setcounter{groupcount}{0}
                        \draw [
                            shorten >=-#5,
                            shorten <=-#5
                        ] (startgroup) -- node [anchor=north] {#3} (endgroup);
                    \fi
                \fi
            }
            \ifnum\thegroupcount=1
                        \setcounter{groupcount}{0}
                        \draw [
                            shorten >=-#5,
                            shorten <=-#5
                        ] (startgroup) -- node [anchor=north] {#3} (endgroup);
            \fi
        }
    }
}

\begin{abstract}

Domain-specific languages that execute image processing pipelines on GPUs, such
as Halide and Forma, operate by 1)~dividing the image into overlapped tiles,
and 2)~fusing loops to improve memory locality. However, current approaches
have limitations: 1)~they require intra thread block synchronization, which has
a nontrivial cost, 2)~they must choose between small tiles that require more
overlapped computations or large tiles that increase shared memory access (and
lowers occupancy), and 3) their autoscheduling algorithms use simplified GPU
models that can result in inefficient global memory accesses.

We present a new approach for executing image processing pipelines on GPUs that
addresses these limitations as follows. 1)~We fuse loops to form overlapped tiles
that fit in a single \emph{warp}, which allows us to use lightweight
warp synchronization. 2)~We introduce \emph{hybrid tiling}, which
stores overlapped regions in a combination of thread-local registers and shared memory.
Thus hybrid tiling either increases occupancy by decreasing shared memory usage 
or decreases overlapping computations using larger tiles. 
3)~We present an automatic loop fusion algorithm that considers several factors
that affect the performance of GPU kernels. We implement these techniques in
PolyMage-GPU, which is a new GPU backend for PolyMage. Our approach produces code that
is faster than Halide's manual schedules: 1.65$\times$ faster on an NVIDIA GTX
1080Ti and 1.33$\times$ faster on an NVIDIA Tesla V100. \end{abstract}

\maketitle

\section{Introduction}
\label{sec:intro}

Image processing programs are essential in several domains, including computer
vision, embedded vision, computational photography, and medical imaging. These
programs run on a variety of platforms, from embedded systems to
high-performance clusters that process large amounts of image data.
With the increasing demand
and sophistication of image processing computations (including real-time
requirements), there is a growing need for high-performance
implementations of image processing programs. 

An image processing program is logically structured as a directed acyclic graph
of connected stages, where each stage performs per-pixel data parallel
operations on its input image and produces an output image for dependent stages.
There
are several domain-specific languages (DSLs) for writing image processing pipelines,
including Halide~\cite{pldi13:halide}, PolyMage~\cite{asplos15:polymage}, and
Forma~\cite{gpgpu8:forma}. These DSLs allow the programmer to write independent
stages in a natural way, but still get high-performance code by applying key
optimizations, including \textit{loop fusion} 
and \emph{overlapped tiling}.
Loop fusion allows the program to exploit locality, and is
performed on the basis of a schedule that is either specified by an
expert~\cite{pldi13:halide,gpgpu8:forma} or automatically generated using
heuristics~\cite{sigraph16:halide-cpu-auto,siggraph19:halide-auto-dl,siggraph18:halide-auto-diff,ppopp18:polymage-dp}.
After loop fusion, overlapped tiling~\cite{asplos15:polymage, pldi13:halide,gpgpu8:forma} 
splits each stage into overlapping regions (known as tiles) that can be processed in parallel
without synchronization with other tiles. On a GPU, each tile is mapped to a
\emph{thread block}, which stores intermediate results (scratchpad arrays) in
shared memory.

These approaches~\cite{ics12:holewinski, taco13:ppcg, pact16:rawat, ppopp17:prajapati,
ics08:baskaran,pldi13:halide,gpgpu8:forma, ipdps19:artemis, taco2019:flextended-tiles} 
to overlapped tiling and automatic loop fusion give suboptimal performance on modern GPUs
for three reasons. 1)~Processing an overlapped tile per
thread block has a high synchronization cost across stages. 2)~Smaller tiles
have more overlapped regions (and thus require more redundant computation),
but larger tiles require more shared memory accesses (and thus lower occupancy).
3)~State-of-the-art autoscheduling algorithms for loop fusion and tile-size
selection do not employ a rich cost model for GPUs. 
For example, cost models in~\cite{siggraph18:halide-auto-diff,pact16:rawat} do not consider the number of global memory transactions,
the ability to hide latency of global memory accesses, and occupancy.


We present PolyMage-GPU (based on PolyMage~\cite{asplos15:polymage}), a
compiler for image processing pipelines that leverages the architecture of
modern GPUs to generates high performance code. PolyMage-GPU exploits the fact
that all threads in a \emph{warp} can synchronize using warp synchronization,
which has significantly lower overhead than thread block synchronization. In
addition, modern GPUs have \textit{warp shuffle}~\cite[Chapter
B.16]{cuda-guide} instructions that allow threads in a warp to read each
others' register values. PolyMage-GPU uses warp shuffles to lower shared
memory usage and support larger overlapped tiles. Finally, we develop a cost
model for GPUs that accounts for several factors, including the number of
global memory transactions, occupancy, and resource utilization. We use this
cost model to determine the optimal tile and thread block sizes and loops to
fuse, using Dynamic Programming Fusion~\cite{ppopp18:polymage-dp}.

To summarize, this paper makes the following contributions.
\begin{enumerate}

  \item We present an approach to overlapped tiling on GPUs that executes one
  overlapped tile per warp, which significantly decreases synchronization costs
  (Section~\ref{sec:otpw}).

  \item We present \emph{hybrid tiling}, which stores portions of a tile in
  registers instead of shared memory, which reduces the fraction of redundant
  computations, and reduces shared memory utilization. This improves performance
  by decreasing global memory loads and increasing occupancy. Hybrid tiling
  relies on \emph{warp shuffle} instructions available in recent GPUs
  (Section~\ref{sec:hybrid-tiling}).
  
  \item We present a fast automatic fusion and tile size selection algorithm
  that considers key factors affecting the performance of an image processing
  pipeline on a GPU, including the number of global memory transactions, 
  fraction of redundant computations, and occupancy (Section~\ref{sec:cost-model}).
  
  \item We implement the aforementioned techniques in PolyMage-GPU, which is a
  new GPU backend for PolyMage~\cite{asplos15:polymage}, which is a DSL
  embedded in Python for writing image processing pipelines.
  
  \item Using established benchmarks, we compare our approach to manually written
  schedules in Halide. On a GeForce GTX 1080Ti, we 
  achieve a speedup of 1.65$\times$ over manual
  schedules and on a Tesla V100, we achieve a speedup of 1.33$\times$ over manual schedules.

\end{enumerate}

The rest of this paper is organized as follows. Section~\ref{sec:background}
discusses the architecture of NVIDIA GPUs, the PolyMage DSL, and Dynamic
Programming Fusion. Section~\ref{sec:overview} presents an overview of our
approach. Section~\ref{sec:otpw} presents our technique for
running one overlapped tile per warp. Section~\ref{sec:hybrid-tiling}
presents hybrid tiling. Section~\ref{sec:cost-model} presents our automatic
fusion algorithm. Section~\ref{sec:impl-eval} evaluates our work.
Section~\ref{sec:related-work} discusses related work. Finally,
Section~\ref{sec:conclusion} concludes.


\section{Background}
\label{sec:background}

This section first presents the essentials of GPU architecture that are
necessary for our work. We then present the PolyMage DSL for writing image
processing programs, and two key ideas that it employs: dependence vectors and
dynamic programming fusion.

\subsection{NVIDIA GPU Architecture}

An NVIDIA GPU consists of several Simultaneous Multiprocessors (SM) that
execute one or more thread blocks. Each SM consists of several CUDA cores,
shared memory, and registers. The number of warps that an SM can execute
concurrently depends on properties of the running CUDA kernel: the number of
thread blocks it has, the number of
threads per thread block, the shared memory used by each thread block, and the
registers used by each thread. The \emph{occupancy} is the ratio of the number of
concurrently executing warps to the maximum number of warps supported. When a
warp accesses global memory, its execution is delayed due to memory access
latency. To hide this latency, the warp scheduler switches execution to another
warp that is ready to execute.

CUDA threads can synchronize in two ways. \emph{Thread block synchronization}
synchronizes all threads in a block: until all warps in the block reach the
same \texttt{\_\_syncthreads} statement, no warp is allowed to proceed.
However, as mentioned above, when a warp is stalled on a global memory access,
the SM tries to run another warp. Thread block synchronization can force an SM to
idle if all warps are waiting for memory accesses to be satisfied.
Contemporary current stencil code generators for GPUs use thread block
synchronization between producer-consumer stages (Section~\ref{sec:related-work}).
In contrast, \emph{warp synchronization} synchronizes all
threads in a warp, and no thread can proceed until all threads in the warp reach the
synchronization point (\texttt{\_\_syncwarp}). However, other warps in the same
thread block can make progress, thus it is more lightweight than thread block
synchronization.

\begin{figure}[t]
  \small
      \begin{lstlisting}[language=C,
      mathescape=true, numbers=left, numberstyle=\footnotesize,
      columns=fixed,
          basicstyle=\ttfamily\footnotesize,
          keywordstyle=\textbf,
          firstnumber=1,
  escapeinside={(*}{*)}, % if you want to add comments in code
  morekeywords={blockDim, blockIdx, threadIdx, warpSize, \_\_shared\_\_},
          escapechar=|, xleftmargin=1.0ex,numbersep=4pt]
int val = rand ();
for (int offset = 16; offset > 0;offset /= 2)
  val += __shfl_sync(0xffffffff, val, 
                     threadIdx.x+offset, warpSize);
      \end{lstlisting}
      \vspace{-1em}
  \caption{\small CUDA kernel invoked with 32 threads in \textit{x}-dimension. At each iteration,
  each thread add next \texttt{offset} thread's \texttt{val} to its \texttt{val}. At the end of 
  loop, \texttt{val} of the first thread contains the sum.}
  \label{fig:shfl-example}
  \vspace{-1.5em}
\Description[]{}
\end{figure}

The \emph{warp shuffle} instructions~\cite[Chapter
B.16]{cuda-guide}\cite{amd-hip-warp-shuffle} available in recent AMD and NVIDIA
GPUs allow threads to read register values from other threads in the same warp.
The \texttt{\_\_shfl\_sync} instruction takes four arguments: a 32-bit mask of
threads participating in the shuffle, the variable stored in the register to
read, the index of the source thread containing the register, and the warp
size. Similarly, \texttt{\_\_shfl\_down\_sync} and \texttt{\_\_shfl\_up\_sync}
read registers from a thread with an index immediately before or after the
calling thread. Figure~\ref{fig:shfl-example} shows an example
from~\cite{cuda-warp-level-primitives} of reduction using
\texttt{\_\_shfl\_sync}. For a shuffle to succeed both the calling thread and
source thread must execute the instruction.

\subsection{PolyMage DSL}

PolyMage~\cite{asplos15:polymage} is a DSL embedded in Python for writing image
processing pipelines. The PolyMage compiler transforms programs in the DSL into
high-performance code for CPUs.
Figure~\ref{fig:polymage-blur} shows
an image blurring program (\textit{blur}) with two stages (\texttt{blurx} and \texttt{blury}).
The parameters to the pipeline are the number of rows and columns in the 
image (line~\ref{line:blur:parameter}). The program first feeds the input image
(\texttt{img} on line~\ref{line:blur:image}) to \texttt{blurx}, and then the output
of \texttt{blurx} to \texttt{blury}.
Each stage is a function mapping a multi-dimensional integer domain to values
representing intensities of image pixels (lines~\ref{line:blur:blurxstart}
and~\ref{line:blur:blurystart}). The domain of the function is defined
at lines~\ref{line:blur:intervalstart}--\ref{line:blur:intervalend}.
\texttt{blurx} takes the image as input and blurs it in the \textit{x}-direction
(lines~\ref{line:blur:blurxstart}--\ref{line:blur:blurxend}).
\texttt{blury} blurs the output of \texttt{blurx} in the \textit{y}-direction and produces final 
output (lines~\ref{line:blur:blurystart}--\ref{line:blur:bluryend}).
The PolyMage compiler performs loop fusion on producer-consumer stages to
improve locality and provide parallel execution. When fusing two stages,
PolyMage performs overlapped tiling using polyhedral transformations. Two adjacent
tiles perform redundant computations to ensure that all the data required to
compute the output of a tile (known as \emph{liveouts}) is available within that tile,
providing parallel execution of all tiles. Within a tile, the output
of a producer stage is transferred to its consumer using small buffers, known as
\textit{scratchpads}. A scratchpad is small enough to fit in a CPU cache, or
in our work,  in GPU shared memory or registers.

\subsection{Dependence Vectors} 
\label{sec:background:dependence-vector}

PolyMage uses dependence vectors to encode the dependencies between consumer
and producer stages. A \emph{dependence vector}~\cite{wolfe-dependence-vector} is the difference of the time
stamps when a value is consumed and when it is produced. For example, in the
\textit{blur} program, the \texttt{blury} stage, at \texttt{(2,c,x,y)},
consumes values that the \texttt{blurx} stage produces at \texttt{(1,c,x,y-1)},
\texttt{(1,c,x,y)}, and \texttt{(1,c,x,y+1)}. This is captured by the
dependence vectors (1,0,0,-1), (1,0,0,0), and (1,0,0,1).

\begin{figure}[!tb]
   \footnotesize
\begin{lstlisting}[language=Python,
        mathescape=true, numbers=left, 
        % numberstyle=\small, 
            basicstyle=\ttfamily,
            firstnumber=1,
            keywordstyle=\textbf,
    escapeinside={(*}{*)}, % if you want to add comments in code
    morekeywords={Domain,Function,Interval,Accumulator,Variable,
          Constant,Parameter,Kernel2D,Correlate,Min,Max,Sum,
          Image,Case,Default,Upsample,Downsample,Replicate,
          Reduce, UChar, Char, Float, Double, Int, UInt, 
          Short, UShort, Long, ULong, Condition, Stencil, Accumulate},
            escapechar=|, xleftmargin=3.0ex,numbersep=4pt]
R,C = Parameter(Int,"R"),Parameter(Int,"C") |\label{line:blur:parameter}|

# Vars
x = Variable(Int,"x")
y = Variable(Int,"y")
c = Variable(Int,"c")

# Input Image
img = Image(Float,"img",[3,R+2,C+2]) |\label{line:blur:image}|

# Intervals
cr = Interval(Int,0,2) |\label{line:blur:intervalstart}|
xrow,xcol = Interval(Int,1,R),Interval(Int,0,C+1) 
yrow,ycol = Interval(Int,1,R),Interval(Int,1,C) |\label{line:blur:intervalend}|

cond = Condition(x,'>=',1) & Condition(x,'<=',R) & 
       Condition(y,'<=',C) & Condition(y,'>=',1)

blurx = Function(([c,x,y],[cr,xrow,xcol]),|\label{line:blur:blurxstart}|
		 Float,"blurx") 
blurx.defn = [Case(cond,(img(c,x-1,y) + img(c,x,y) +
                         img(c,x+1,y))/3)] |\label{line:blur:blurxend}|
blury = Function(([c,x,y],[cr,yrow,ycol]),
		 Float,"blury") |\label{line:blur:blurystart}|
blury.defn = [Case(cond,(blurx(c,x,y-1) + 
              blurx(c,x,y) + blurx(c,x,y+1))/3)] |\label{line:blur:bluryend}|
\end{lstlisting}
    \vspace{-1em}
    \caption{\small PolyMage DSL specification for \textit{blur}.
    \vspace{-1em}
    }
    \label{fig:polymage-blur}
\Description[]{}
\end{figure}

\subsection{Dynamic Programming Fusion}

\emph{Dynamic Programming Fusion} (DP-Fusion)~\cite{ppopp18:polymage-dp} is an
algorithm that performs automatic fusion of image processing pipelines in a few
seconds. DP-Fusion finds schedules that are competitive with take days for an
autotuner, and are better than a greedy CPU
autoscheduler~\cite{sigraph16:halide-cpu-auto}. Instead of using a greedy
algorithm and a simple cost function, DP-Fusion enumerates all valid fusion
possibilities and uses dynamic programming combined with an analytic cost
function to significantly decrease the runtime of a combinatorial algorithm.
Among all fusion possibilities, DP-Fusion finds the best fusion choices on the
basis of the cost of candidate fused loops. The cost of fused loops is
calculated using a cost function that also uses a model to determine tile
sizes. 
PolyMage uses DP-Fusion to find the best schedules for image processing programs executing on multi-core CPUs~\cite{ppopp18:polymage-dp}.
In this paper, we present a cost model for GPUs that integrates with DP-Fusion.

\section{Overview}

\label{sec:overview}
\begin{figure}[!tb]
\footnotesize
\begin{lstlisting}[language=C,
      mathescape=true, numbers=left, numberstyle=\text, 
          basicstyle=\ttfamily,
          keywordstyle=\textbf,
          firstnumber=1,
  escapeinside={(*}{*)}, % if you want to add comments in code
  morekeywords={blockDim, blockIdx, threadIdx, warpSize, shared},
          escapechar=|,xleftmargin=3.0ex,numbersep=4pt]
blur_otptb(img[3][R][C], blury[3][R-2][C-2])
  shared blurx[blockDim.y][tile*blockDim.x+2];|\label{line:optpb:shared-mem}|
  c = threadIdx.z
  y = blockIdx.y*blockDim.y + threadIdx.y
  for (tx = 0; tx < tile+1; tx++) |\label{line:optpb:blurx-loop-starts}|
    xx = tx * blockDim.x + threadIdx.x
    x = (blockIdx.x*blockDim.x)*tx + xx
    if (xx < tile*blockDim.x+2)
      blurx[y][xx] = (img[c][y-1][x]+img[c][y][x]+
		      img[c][y+1][x])/3 |\label{line:optpb:blurx-loop-ends}|
  __syncthreads ();|\label{line:optpb:syncthread}|
  for (tx = 0; tx < tile; tx++) |\label{line:optpb:blury-loop-starts}|
    xx = tx * blockDim.x + threadIdx.x
    x = (blockIdx.x*blockDim.x)*tx + xx
    blury[c][y][x] = (blurx[y][xx-1]+blurx[y][xx]+
		      blurx[y][xx+1])/3  |\label{line:optpb:blury-loop-ends}|
\end{lstlisting}

  \caption{\small Equivalent CUDA code generated by Halide for \textit{blur}. 
  Both \textit{blurx} and \textit{blury} are fused in an overlapped tile of size \texttt{tile} in $x$ and 1 in $y$, which is computed by one 
  thread block. 
  \vspace{-1em}
  }
  \label{fig:cuda-blur-otptb}
\Description[]{}
\end{figure}




\begin{figure}[t]
  \small
  \includegraphics[width=\linewidth]{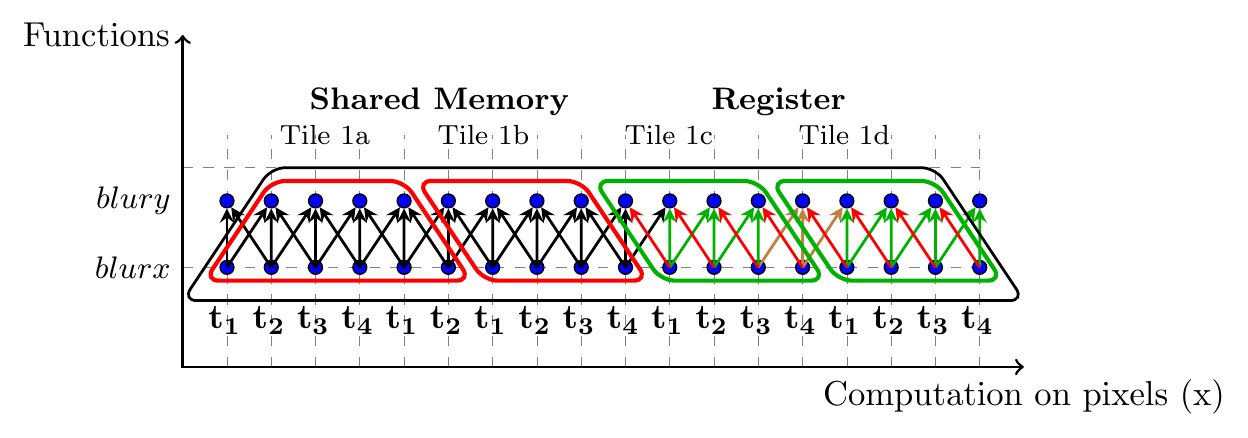}
  \caption{
    \small Hybrid Tiling for \textit{blur} program with tile size of 2 in x-dimension and warp size of 4. 
  The overlapped tile is split into four tiles. Tiles in {\color{red}{red}}
  are stored in shared memory and tiles in {\color{green!70!black}{green}} are stored in registers.
  Each point of \texttt{blurx} is computed and stored in the register of same thread represented by t$_\text{i}$. 
  The producer loads in {\color{red}{red}} are from the registers of current thread (Type \textcircled{1}), {\color{black}{black}} are from shared memory (Type \textcircled{2}), 
  {\color{green!70!black}{green}} are from the registers of another thread in same parallelogram tile (Type \textcircled{3}), and {\color{brown}{brown}} are from registers of another thread in previous parallelogram tile (Type \textcircled{4}). \label{fig:typ-producer-loads} \label{fig:regtiling} 
  \vspace{-2.5em}
  }
\Description[]{}
\end{figure}
 
\begin{figure}[!tb]
\footnotesize
\begin{lstlisting}[language=C,
      mathescape=true, numbers=left, firstnumber=1, numberstyle=\footnotesize, 
          basicstyle=\ttfamily,
          keywordstyle=\textbf,
          commentstyle=\textit,
  escapeinside={(*@}{@*)}, % if you want to add comments in code
  morekeywords={blockDim, blockIdx, threadIdx, warpSize, shared},
          escapechar=|,xleftmargin=3.0ex,numbersep=4pt]
blur_otpw_ht(img[3][R][C], blury[3][R-2][C-2])
  shared blurx[blockDim.y][blockDim.x/warpSz]
              [tile/2*warpSz+2];
  y = blockIdx.y * blockDim.y+threadIdx.y;
  c = threadIdx.z;
  warpSz = warpSize;
  warp = threadIdx.x/warpSz;
   
  for(tx = 0; tx < 2; tx++) |\label{line:otpw-hybrid:blurx-shared-start}|
   for(txx = tx*tile/4;txx<(tx+1)*tile/4+1;txx++)
    xx = tile*warpSz+threadIdx.x%warpSz;
    x = (blockIdx.x+1)*blockDim.x*tx+threadIdx.x;
    if(xx < tile*warpSz+2)
     blurx[y][warp][xx] = (img[c][y-1][x]+
     img[c][y][x]+img[c][y+1][x])/3;|\label{line:otpw-hybrid:blurx-shared-end}|
  x = warp_idx+8*warpSz+lane_id_x; |\label{line:otpw-hybrid:blurx-reg-8}|
  blurx_8 = (img[c][y-1][x]+img[c][y][x] +
             img[c][y+1][x])/3;
  x = warp_idx+9*warpSz+lane_id_x;
  blurx_9 = (img[c][y-1][x]+img[c][y][x]+
             img[c][y+1][x])/3; 
  /*similarly, for all iterations till 15*/|\label{line:otpw-hybrid:blurx-reg-15}|
  syncwarp();
  for(tx = 0; tx < 2; tx++)|\label{line:otpw-hybrid:blury-shared-start}|
   for(txx = tx*tile/4; txx <(tx+1)*tile/4; txx++)
    xx = tile*warpSz+threadIdx.x%warpSz;
    x = (blockIdx.x+1)*blockDim.x*tx+threadIdx.x;
    if(xx > 0 and xx < tile/2*warpSz+2)
     blury[c][y][x] = (blurx[y][warp][xx-1]+
      blurx[y][warp][xx]+blurx[y][warp][xx+1])/3;|\label{line:otpw-hybrid:blury-shared-end}|
  blurx_l_2_8 = shfl_up(FULL_MASK, blurx_8, 2);  |\label{line:otpw:blury-reg-blurx_l_2_8}|
  blurx_l_1_8 = shfl_up(FULL_MASK, blurx_8, 1);  |\label{line:otpw:blury-reg-blurx_l_1_8}|
  if(lane_id_x == 0) |\label{line:otpw:blury-reg-blurx_8_sh_mem-start}|
   blurx_l_2_8=blurx[y][warp][7*warpSz+warpSz-2];
   blurx_l_1_8=blurx[y][warp][7*warpSz+warpSz-1];
  if(lane_id_x == 1) 
   blurx_l_2_8=blurx[y][warp][7*warpSz+warpSz-1];|\label{line:otpw:blury-reg-blurx_8_sh_mem-end}|
  x = warp_idx+8*warpSz+lane_id_x;
  blury[c][y][x] = (blurx_l_2_8+blurx_l_1_8+
                    blurx_8)/3;|\label{line:otpw:blury-reg-8}|
  blurx_l_2_9 = shfl_up(FULL_MASK, blurx_9, 2); |\label{line:otpw:blury-reg-9-prev-start}|
  blurx_l_1_9 = shfl_up(FULL_MASK, blurx_9, 1);
  _blurx_l_2_9=shfl(FULL_MASK, blurx_8, warpSz-2);
  _blurx_l_1_9=shfl(FULL_MASK, blurx_8, warpSz-1);
  if(lane_id_x == 0) 
    blurx_l_1_9 = _blurx_l_1_9;
    blurx_l_2_9 = _blurx_l_2_9;
  if(lane_id_x == 1) blurx_l_2_9 = _blurx_l_1_9; |\label{line:otpw:blury-reg-9-prev-end}|
  x = warp_idx+9*warpSz+lane_id_x;  
  blury[c][y][x] = (blurx_l_2_9+blurx_l_1_9+
                    blurx_9)/3; |\label{line:otpw:blury-reg-9}|
  /*similarly, for all iterations till 15*/
\end{lstlisting}

  \caption{\small CUDA code for \textit{blur}, with \textit{blurx} and \textit{blury} fused in
  an overlapped tile of size \texttt{tile} in the \textit{x}-dimension, which is computed by one warp.
   The first half of  the tile is stored in shared memory with latter half in
  registers. In this code \texttt{shfl*} refers to \texttt{\_\_shfl*\_sync}.
  \vspace{-1.5em}
  }
  \label{fig:cuda-blur-reg-tile}
\Description[]{}
\end{figure}

Figure~\ref{fig:cuda-blur-otptb} shows CUDA code that is equivalent to the code
that Halide produces for \textit{blur}. The code fuses both \textit{blurx} and
\textit{blury} together and uses overlapped tiles of length \texttt{tile} in
the \textit{x}-dimension and unit length in \textit{y}-dimension. During
execution, all threads in a thread block 1)~compute \textit{blurx} in parallel
by looping over all points in the tile
(lines~\ref{line:optpb:blurx-loop-starts}--\ref{line:optpb:blurx-loop-ends}),
2)~store the result of \textit{blurx} in a scratchpad (which is in shared
memory), 3)~use thread-block synchronization to ensure that all \textit{blurx}
values are ready (line~\ref{line:optpb:syncthread}), and 4)~calculates
\textit{blury} in parallel, which depends on \textit{blurx}
(line~\ref{line:optpb:blury-loop-starts}--~\ref{line:optpb:blury-loop-ends}).
On an NVIDIA GTX 1080Ti, this code exhibits its best performance (1.40ms) on a
$4096\times$$4096\times$$3$ input with 8 tiles and block sizes of
$64\times$4$\times$1. However, thread block synchronization can lower
occupancy, so there is room for improvement.

\paragraph{Overlap Tile per Warp (OTPW)} We modify the program to assign each
overlapped tile to a warp, instead of a thread block. This change allows us to
use warp synchronization (\texttt{\_\_syncwarp}), which allows the SM to
execute a one warp even if another warp is waiting for a memory access. This
code exhibits its best performance (1.35ms) with 8 tiles and block sizes of
$64\times$$4\times$$1$. This is a 1.04$\times$ speedup over the prior approach.
This choice of tile size produces 0.8\% redundant computations per warp. We can
achieve fewer redundant computations (0.4\%) with tile size 16, but that increases
running time (1.45ms) because it consumes far more shared memory (over 16KB). This limits
the number of warps that the GPU can run concurrently, i.e., occupancy is only
62.5\%.


\paragraph{Hybrid Tiling}

To further improve performance, we introduce \emph{hybrid tiling}, which is a
technique that decreases the size of the scratchpad buffer in shared memory, by
storing some parts of the overlapped tile in registers. In the earlier
approaches, we employed the scratchpad to share values between consumers
(\textit{blury}) and producers (\textit{blurx}) in a thread block. However,
since we now assign each tile to a warp, we can use \emph{warp shuffle}
instructions that allow threads in a warp to read register values from other
threads in the same warp. This eliminates the need for per thread redundant
computation that arise in register blocking.
Figure~\ref{fig:regtiling} sketches the structure of the computation,
assuming four tiles: the first two tiles are stored in shared memory, whereas
the latter two tiles are stored in registers. When a \textit{blury} value
depends on a \textit{blurx}-value in a register, it can read it directly,
using warp shuffles to read across threads if needed.


On a GTX 1080Ti, the code so far only uses 24 registers. With a tile size of
16, we can store half of the tile in registers, which halves the shared memory
usage, and leads to 100\% occupancy. With hybrid tiling, the code runs in 1.2ms
which is 1.13$\times$ faster than the \textit{OTPW} approach, and 1.16$\times$
faster than the original program.

Figure~\ref{fig:cuda-blur-reg-tile} sketches the CUDA code for \textit{blur}
that uses \emph{overlap tile per warp} and \emph{hybrid tiling}. In the figure,
the data points of \textit{blurx} for first two tiles are stored in shared
memory while the later tiles are stored in the registers.
Lines~\ref{line:otpw-hybrid:blurx-shared-start}--\ref{line:otpw-hybrid:blurx-shared-end}
processes \textit{blurx} on the first two tiles stored in shared memory using a
warp by assigning consecutive data points to consecutive threads in a warp and
looping over all points in both tiles.
Lines~\ref{line:otpw-hybrid:blurx-reg-8}--\ref{line:otpw-hybrid:blurx-reg-15}
unroll the loop and store each data point in registers for two register tiles.
Lines~\ref{line:otpw-hybrid:blury-shared-start}--\ref{line:otpw-hybrid:blury-shared-end}
compute the values of \textit{blury} for first two tiles that are stored in
shared memory.
Lines~\ref{line:otpw:blury-reg-blurx_l_2_8}--\ref{line:otpw:blury-reg-blurx_l_1_8}
retrieve the values of \textit{blurx} from other threads using \textit{warp
shuffle}. Since the first two values for the first thread in a warp are the
values produced and stored in shared memory by last two threads of that warp,
lines~\ref{line:otpw:blury-reg-blurx_8_sh_mem-start}--\ref{line:otpw:blury-reg-blurx_8_sh_mem-end}
retrieve the last two values of shared memory for that warp.
Line~\ref{line:otpw:blury-reg-8} computes each \textit{blury} point for the
eighth iteration of the larger overlapped tile. Similarly, for the ninth
iteration,
lines~\ref{line:otpw:blury-reg-9-prev-start}--\ref{line:otpw:blury-reg-9-prev-end}
retrieve the values of \texttt{blurx\_9} from previous threads and for first
two threads of warp values of \texttt{blurx\_8} are retrieved from last two
threads of the warp. We generate code for the remaining six iterations in the
same manner.

\paragraph{Loop Fusion}

The final problem involves choosing tile and block sizes. We present an
automatic fusion algorithm that considers key factors affecting the performance
of GPU kernels which are not considered in previous
work~\cite{siggraph19:halide-auto-dl,sigraph16:halide-cpu-auto,siggraph18:halide-auto-diff}:
1)~number of global memory transactions, 2)~achieved and theoretical
occupancy, 3)~GPU resource usage, and 4)~fraction of overlapping
computations.
\medskip

\tikzstyle{block} = [rectangle, draw, fill=blue!20, 
    text width=6em, text centered, rounded corners, minimum height=4em]
\tikzstyle{newblock} = [rectangle, draw, fill=red!20, 
    text width=6em, text centered, dashed, rounded corners, minimum height=4em]
\tikzstyle{line} = [draw, -latex']

\begin{figure*}
  \small
\begin{tikzpicture}[node distance = 2cm, auto]
    \node [block] (dsl-spec) {DSL Spec};
    \node [block, right of=dsl-spec, node distance=3cm] (init) {Create DAG\\Polyhedral Representation };
    \node [newblock, right of=init, node distance=3cm] (fuse) {Automatic Fusion for GPUs};
    \node [newblock, right of=fuse, node distance=3cm] (otpw) {Overlapped Tile Per Warp};
    \node [newblock, right of=otpw, node distance=3cm] (ht) {Hybrid Tiling};
    \node [block, right of=ht, node distance=3cm] (code-gen) {CUDA Code Generation};

    \path [line] (dsl-spec) -- (init);
    \path [line] (init) -- (fuse);
    \path [line] (fuse) -- (otpw);
    \path [line] (otpw) -- (ht);
    \path [line] (ht) -- (code-gen);
\end{tikzpicture}
\caption{Compilation pipeline of image processing program written in
PolyMage-GPU, which is based on PolyMage~\cite{asplos15:polymage}. The three
phases in middle with dashed rectangles are the new phases of
PolyMage-GPU (Sections~\ref{sec:otpw}--\ref{sec:cost-model}).}
\label{fig:pipeline}
\vspace{-1em}
\Description[]{}
\end{figure*}
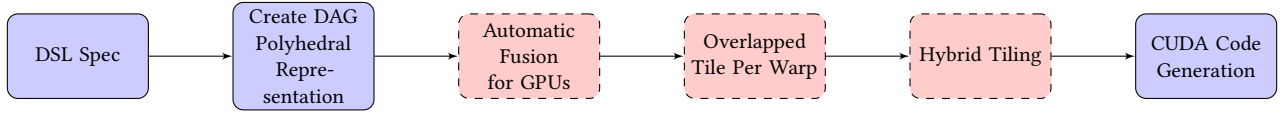

We implement OTPW, hybrid tiling, and our new new fusion algorithm in
PolyMage-GPU. Figure~\ref{fig:pipeline} shows the structure of the compilation
pipeline. In summary, our approach uses low cost synchronization, distributes
tile in shared memory and registers, decreases shared memory usage, and enables
larger tiles to decrease number of overlapping computations without any loss in
occupancy. We also address the problem of fusing pipeline stages and choosing
tile and thread block sizes automatically.

\section{Overlap Tile per Warp}
\label{sec:otpw}

In this section, we describe how we calculate 1)~the tile size for
each stage, 2)~the assignment of input data points to threads, 3)~the size
of the output scratchpad, and 4)~the fraction of overlap.

Let $(b_x, b_y, b_z)$ be the coordinates of a thread block
$(B_x, B_y, B_z)$ be the thread block size. Consider a group of fused stages with tile sizes
$(T_x, T_y, T_z)$ that consumes a three-dimensional input of size $(N_x, N_y,
N_z)$, where each dimension is labelled $i \in \{x, y, z\}$. We convert the
three-dimensional coordinates of a thread $(t_x, t_y, t_z)$ to
 a linear thread ID: $t_x+B_x\times t_y+B_x\times
B_y\times t_z$. The Warp ID of a thread is the thread ID divided by \warpSize{}
and the index of a thread in a warp (known as its \emph{lane ID}) of a thread
is the remainder. We define warp sizes, $W_x$, $W_y$, $W_z$
such that:

\begin{eqnarray*}
W_x & = &\mathtt{minimum } (B_x, \texttt{WarpSize})\\
W_y & = &\mathtt{minimum } (B_y, \texttt{WarpSize}\div{W_x})\\
W_z & = &\mathtt{minimum } (B_z, \texttt{WarpSize}\div{(W_x \times W_y)})
\end{eqnarray*}

In these equations we assume that number of threads in a thread block
are a multiple of \warpSize{}. (We add extra threads as padding if needed.)
These warp sizes are the number of threads with distinct IDs of that dimension in a 
warp. The number of warps in dimension $i$ in a thread block is equal to the ratio of block 
size to the warp size of that dimension
($\left \lceil B_i / W_i \right \rceil$). The warp ID of a thread in a 
dimension is 
the floor of division of the thread's ID in that dimension to the warp size 
of that dimension, i.e.($\left \lfloor (b_i \times B_i + t_i) / W_i \right\rfloor$).
Moreover, the lane ID is the remainder ($(b_i \times B_i + t_i) \bmod W_i$).
 Note that product of all the warp sizes obtained using these
equations is equal to \texttt{WarpSize}.
For given overlapped tile sizes, we create a \emph{warp overlapped tile} by
extending the tile sizes of each dimension to cover exactly one warp. The total 
number of points in a warp overlapped tile excluding the redundant 
computations is the product of the number of points in the given overlapped tile sizes 
and \warpSize. For the given overlapped tile size, the size of the warp 
overlapped tile is $(T_x\times W_x, T_y\times W_y, T_z \times W_z)$. 
For example, if the tile size is (8,4,1), block size is (16,8,1), then the warp size will be (16,2,1) 
and the warp overlapped tile size will be (128,8,1). 

Tiling a dimension produces two dimensions: an outer dimension that is iterated from
the number of tiles and an inner dimension that is iterated  tile size times.
We initialize the outer dimension to the warp ID of that dimension, and the
inner dimension to the sum of the lane ID and the 
product of current tile iteration and \warpSize{}. 
To process each warp tile, we assign consecutive threads in the $i^{\text{th}}$ dimension
to consecutive data points in an outer loop that runs for $T_i$ times. 

The size of each scratchpad for a stage is exactly the number of data points 
computed by the thread block for that stage.
For the $n^{\text{th}}$ stage, each warp computes two types of data points in
 $i^{\text{th}}$ dimension: 1)~$T_i \times W_i$ computations for the tile, and 
 2)~$O_i^{n}$ overlapping computations. 
We represent the number of data points computed (and the size of the scratchpad) for 
$n^{\text{th}}$ stage  as $\prod_{i \in \{x, y, z\}} \left \lceil B_i / W_i \right \rceil \times (T_i \times W_i + O_i^{n})$. 

Since tiling introduces extra conditional branches and
arithmetic instructions, we do not perform \text{OTPW} in a dimension when
the warp size in that dimension is 1. 
However, as long as the group of stages processes more than one input point, at least
one dimension will have warp size greater than 1.

\section{Hybrid Tiling}
\label{sec:hybrid-tiling}

In this section we present \emph{hybrid tiling}, which divides a tile between
shared memory and registers. Hybrid tiling riles on the fact that each
overlapped tile fits in a single wrap. We use \emph{warp shuffle} instructions
to allow each thread to access data from other threads in a warp, which
eliminates the need for certain redundant computations per thread. Hybrid
tiling solves the issues of shared memory only tiling by 1)~storing a part of a
tile in registers to decrease allocated shared memory, 2)~providing extra
storage for larger tile sizes, which results in fewer redundant computations,
and in turn, fewer global memory loads and total computations; and 3)~storing
tiles partially in registers, which leads to faster access to data points.


We split the warp overlapped tile over a \emph{split dimension}, into several
parallelogram tiles with left tiles stored in shared memory and right tiles
stored in registers (Figure~\ref{fig:regtiling}). These smaller parallelogram
tiles are of warp size in the split dimension, and the same size as the warp
overlapped tile in other dimensions. The slope of the parallelogram tiles are
parallel to the right hyperplane of the warp overlapped tile in the split
dimension, which ensures there is no cyclic dependence between two adjacent
tiles. The left parallelogram tiles, including the overlap on the left side, are
stored in shared memory.
Since the right tiles depend on left tiles, we must process the left tiles first.

Since all producer loads by \textit{OTPW} are in the shared memory, we need to convert
these loads to access data stored in registers if necessary.
Figure~\ref{fig:typ-producer-loads} shows that there are four types of producer load:
\textcircled{1} is a load from a register of the current thread, if the load index is same as 
the iteration in the split dimension; \textcircled{2} is a load from shared memory, 
if the load index is less than the lower bound of the register tile in the split dimension;
\textcircled{3} is a load from another thread's register in same tile, 
if the load index in the split dimension is less than the iteration in the split dimension;
and \textcircled{4} is a load from another thread's register from the previous tile,
if the difference between the lane ID of the current thread in the split dimension and 
the difference between the iteration and load index in the split dimension is less than zero.


We now present the
code generation algorithm that uses dependence vectors between producer and
consumer stages. Before executing the hybrid tiling algorithm, we use
PolyMage's alignment and scaling to make the dependence vectors between each
producer-consumer pair constant. Algorithm~\ref{algo:hybrid-tiling} is our
hybrid tiling algorithm. For simplicity, we present the algorithm making two
assumptions. First, we assume that the $x$-dimension is the split dimension.
Second, we assume that the difference between any two dependence vectors in the
same dimension after alignment and scaling is less than the warp size. Several
of our benchmarks satisfy these assumptions. However, it is straightforward to
generalize the algorithm~\cite[Appendix A]{jangda2020modelbased}.

The arguments to the \textsc{2-D-HybridTiling} function are the group of stages ($G$),
tile sizes ($T_x \times T_y$), warp sizes ($W_x\times W_y$), and register tile
size ($\mathit{fracReg}$) as a fraction of the tile size in the split
dimension. The result of the function is CUDA code that does hybrid tiling.
First, the algorithm finds a split dimension with tile size greater than 1
(line~\ref{line:algo:hybrid-tiling:split-dim}). If no such dimension is found,
then tiles must be stored entirely in the shared memory. The rest of the
algorithm assumes that the $x$-dimension is the split dimension. Let
$\phi_{rx}$ and $\phi_{ry}$ be the right hyperplanes of warp overlapped tiles
of $G$ in the $x$ and $y$ dimensions respectively. We first generate the shared
memory tile using the PolyMage compiler, and then generate register
tiles using the \textsc{GenRegTile} function that takes a stage of the group ($H$),
the hyperplanes ($\phi_{rx}$, $\phi_{ry}$), the register tile size ($R_x\times R_y$),
 and the warp sizes ($W_x\times W_y$) as arguments
(lines~\ref{line:algo:hybrid-tiling:gen-shared-mem-tile}--\ref{line:algo:hybrid-tiling:gen-reg-tile}).

For all the iterations in the register tile, including the overlapping
computations, we store each computed value of stage $H$ in a distinct variable,
instead of shared memory (line~\ref{line:algo:hybrid-tiling:store-to-var}). We
replace each producer load in the loop is replaced with either a shared memory
read or a warp shuffle
(lines~\ref{line:algo:hybrid-tiling:update-producer-load-start}--\ref{line:algo:hybrid-tiling:update-producer-load-end}).
We get the dependence vector between the producer and consumer
(line~\ref{line:algo:hybrid-tiling:phi_x}) as $\phi_x$ and $\phi_y$. (Note that
$\phi_x \leq \phi_{rx}$ and $\phi_y \leq \phi_{ry}$ due to overlapped tiling
algorithm.) Figure~\ref{fig:ht-code-cases} shows the code generated for three
cases that arise when generating code for a load \texttt{P[a*x+b][c*y+d]}. The
figure shows two types of source lane IDs that contain the register, which
stores the value of the producer load: 1)~\texttt{currTileSrcLane} is the lane
ID for a source thread in the current parallelogram tile and
2)~\texttt{prevTileSrcLane} is the lane ID for a source thread in the previous
parallelogram tile. Value of both ids in $x$-dimension depends on $\phi_x -
\phi_{rx}$ and in $y$-dimension depends on $\phi_y - \phi_{ry}$. We now explain
each of the three cases in detail. 1)~If $\phi_x = \phi_{rx}$, then the value
needed for this load is stored by the current thread's register and we generate
the code for Type \textcircled{1} (line~\ref{line:algo:hybrid-tiling:case-1}).
2)~When $\phi_x - \phi_{rx} \neq 0$ and the iteration in split dimension, i.e.,
$x$-dimension is first iteration of the register tile, then first $|\phi_x$ -
$\phi_{rx}|$ threads of warp loads from shared memory (Type \textcircled{2})
and remaining threads loads from registers of threads in same parallelogram
tile (Type \textcircled{3}). Figure~\ref{fig:case-2-code} shows the code
generated for this case. The conditional determines whether to load from shared
memory or from another thread's register. The \texttt{\_\_shfl\_sync} function
loads the value from the source thread's register. The function
\texttt{getMask} retrieves the mask of threads that can participate in the warp
shuffle. 3)~Otherwise, if a thread needs to load from another thread's register
that stores value of either the current parallelogram tile (Type
\textcircled{3}) or the previous parallelogram tile (Type \textcircled{4}),
then we generate the code in Figure~\ref{fig:case-3-code}. Two warp shuffles
are generated that are executed by all threads and a conditional expression
selects which loaded value to use.

Instead of generating register array, PolyMage-GPU generates a register access
by computing the value of \texttt{a*x+b} and \texttt{c*y+d} for given register
tile iteration \{x,y\} and converts these values to strings. Hence, it produces
explicit variable names for each element of the register array.

Finally, PolyMage-GPU prevents out of bounds accesses in hybrid tiling in two
ways. First, the image sizes in the generated CUDA code are treated as parameters
that are passed to each CUDA kernel. Hence, the bounds of each stage and the
number of tiles depends on the image sizes. Second, before computation of every
iteration of each stage, PolyMage's compiler adds conditionals to ensure that
for the given image sizes, the tile lies within the correct bounds of current
stage. These conditionals will prevent out of bounds accesses if the generated
code is used for different image sizes.

\paragraph{Register Blocking} Register blocking~\cite{wolfe-tiling} is a well-known technique that stores tile
in registers of parallel threads. However, it generates one overlapped tile per
thread, leading to redundant computations between all threads. In contrast,
Hybrid Tiling eliminates these redundant computations by utilizing warp
synchronous behavior of threads and warp shuffles to access shared memory and
the registers of another thread.

      

\begin{figure}
  \footnotesize
\begin{subfigure}[t]{\linewidth}
    \begin{lstlisting}[language=C,escapechar=|,basicstyle=\ttfamily]
val = Reg_P[x][c*y + d]
  \end{lstlisting}
  \caption{Code for a register access from same thread is generated when the source lane ID is the current lane ID, i.e. 
  $\phi_x = \phi_{rx}$ (Type \textcircled{1}). \label{fig:case-1-code}}
\end{subfigure}

\begin{subfigure}[t]{\linewidth}
  \begin{lstlisting}[language=C,escapechar=|,basicstyle=\ttfamily]
currTileSrcLane = (laneId.x + diffPhi.x) +
  (laneId.y + diffPhi.y)*warpSize.x;
/*Type 3:*/ val = __shfl_sync(getMask(), 
            Reg_P[x][c*y+d], currTileSrcLane);
if (laneId.x + diffPhi.x < 0)  
  /*Type 2:*/ val = ShMem_P[a*x+b][c*y+d];
\end{lstlisting}
\caption{Code generated when current iteration is the first iteration of register tile and $\phi_x - \phi_{rx} \neq 0$. When the sum of lane index and $\phi_x - \phi_{rx}$ is less than zero, then value is accessed from shared memory tile (Type \textcircled{2}), otherwise value is accessed from register of thread in the same parallelogram tile (Type \textcircled{3}).\label{fig:case-2-code}}
\end{subfigure}

\begin{subfigure}[t]{\linewidth}
  \begin{lstlisting}[language=C,escapechar=|,basicstyle=\ttfamily]
prevTileSrcLane = (warpSize.x - 1 + diffPhi.x)+
  (warpSize.y - 1 + diffPhi.y)*warpSize.x;
currTileSrcLane = (laneId.x + diffPhi.x) +
  (laneId.y + diffPhi.y)*warpSize.x;
/*Type 3:*/ val = __shfl_sync(getMask(),
              Reg_P[x][c*y+d], currTileSrcLane);
if (laneId.x + diffPhi.x < 0)
  /*Type 4:*/ val = __shfl_sync(getMask(), 
              Reg_P[x-1][c*y+d], prevTileSrcLane);
\end{lstlisting}
\caption{Code generated when current iteration is not the first iteration of register tile and $\phi_x-\phi_{rx}\neq 0$. When the sum of lane index and $\phi_x-\phi_{rx}$ is less than zero, then value is accessed from  register of last $|\phi_x - \phi_{rx}|$ threads of previous parallelogram tile (Type \textcircled{4}) otherwise value is accessed from register of thread in the same parallelogram tile (Type \textcircled{3}). \label{fig:case-3-code}}
\end{subfigure}

\caption{\small Three code generation cases for a producer \texttt{p[a*x+b][c*y+d]} at iteration \{\texttt{x}, \texttt{y}\} of register tile that generates all four load types of Figure~\ref{fig:regtiling}. Each \texttt{p[a*x+b][c*y+d]} of register tile is replaced with \texttt{val} and one of the above the code is added. \texttt{Reg\_P} is the register array storing register tile of \texttt{p}. \texttt{laneId.x} and \texttt{laneId.y} are the lane indices in x and y dimensions of the current thread. \texttt{warpSize.x} and \texttt{warpSize.y} are the warp sizes in x and y dimensions. \texttt{diffPhi.x} is the value of $\phi_x - \phi_{rx}$. \texttt{diffPhi.y} is the value of $\phi_y - \phi_{ry}$. 
\label{fig:ht-code-cases}
\vspace{-1.5em}
}
\Description[]{}
\end{figure}

\begin{algorithm}[tbp]
  \small
  \caption{Hybrid Tiling}\label{algo:hybrid-tiling}
  \begin{algorithmic}[1]

      \Function{GenRegTile} {H, $\phi_{rx}$, $\phi_{ry}$, R$_x \times$ R$_y$, W$_x \times $W$_y$}
          \ForAll {\{\texttt{x}, \texttt{y}\} $\in$ [1, \ldots R$_x$] $\times$ [1, \ldots R$_y$]}
            \State{Let iteration \{\texttt{x}, \texttt{y}\} be}
            \State{\texttt{H[x][y] = f(P[a*x+b][c*y+d], $\ldots$)}}
            \State{Store \texttt{H[x][y]} in a register array \texttt{Reg\_H[x][y]}} \label{line:algo:hybrid-tiling:store-to-var}
            \ForAll{loads \texttt{P[a*x+b][c*y+d]} $\in$ \texttt{f}} \label{line:algo:hybrid-tiling:update-producer-load-start}
              \State{$\phi_x,\phi_y$ = dependence vectors between \texttt{P[a*x+b][c*y+d]} and \texttt{H[x][y]}}\label{line:algo:hybrid-tiling:phi_x}
              \If{$\phi_x$ == $\phi_{rx}$}
                \State{Generate Type\textcircled{1} code in Figure~\ref{fig:case-1-code}} \label{line:algo:hybrid-tiling:case-1}
              \ElsIf {x == 1}
                \State{Generate Type \textcircled{2} and \textcircled{3} code from Figure~\ref{fig:case-2-code}}
              \Else 
                \State{Generate Type \textcircled{3} and \textcircled{4} code from Figure~\ref{fig:case-3-code}}
              \EndIf
            \EndFor\label{line:algo:hybrid-tiling:update-producer-load-end}
          \EndFor
      \EndFunction
      \Function{2-D-HybridTiling} {G, fracReg, T$_x \times$ T$_y$, W$_x \times $W$_y$}
        \State{splitDim = \textit{a dimension with tile size greater than 1}} \label{line:algo:hybrid-tiling:split-dim}
        \State{\textit{If no split dimension exists then} \textbf{return}}
        \State{\textit{Let $\phi_{rx}$ and $\phi_{ry}$ be right hyperplanes of G in $x$ and $y$}}
        \State{\textit{Let} splitDim is the \textit{$x$-dimension.}}
        \State{\textit{Create parallelogram tiles in $x$-dim of size W$_x$ parallel to $\phi_{rx}$}}
        \State{R$_x$ $\gets$ T$_x \times $ fracReg, S$_x$ $\gets$ T$_x \times $ (1 - fracReg)}
        \State{R$_y$  $\gets$ S$_y$ $\gets$ T$_y$}
        \ForAll {H $\in$ G}
          \State{\textit{Gen. Shared Mem Tile with tile size S$_x \times $S$_y$}} \label{line:algo:hybrid-tiling:gen-shared-mem-tile}
          \State {\textsc{GenRegTile}(H, $\phi_{rx}$, $\phi_{ry}$, R$_x \times$ R$_y$, W$_x \times $W$_y$)} \label{line:algo:hybrid-tiling:gen-reg-tile}
        \EndFor
      \EndFunction
  \end{algorithmic}
\end{algorithm}

\section{Automatic Fusion for GPUs}
\label{sec:cost-model}

In this section, we present an automatic fusion algorithm that selects 1)~sets
of stages to fuse, 2)~their tile sizes, and 3)~their thread block sizes.
Our approach leverages DP-Fusion~\cite{ppopp18:polymage-dp}, which
is an algorithm that efficiently enumerates all fusion possibilities, given
a cost function. We introduce a cost function that calculates the minimum
cost of a sequence of fused loops, along with optimal tile sizes and thread
block sizes.


The inputs to our algorithm include the register usage and running time of
each stage, prior to fusion. We gather this information by generating code
for each individual stage, where global memory loads are replaced
with shared memory loads, loops perform a single iteration, and the outermost
loop is nested inside a loop with a large number of iterations (e.g., one
million), to ensure that time measurements are correct.
We obtain the time for each  iteration  by measuring the time taken to execute
the kernel by one thread block, with one thread and divide this by the number
of loop iterations. We measure the register usage of each stage with \texttt{nvcc}.

\begin{table}[t]

\begin{tabular}{|l|c|c|}
  \hline
  \thead{Model} & \thead{GTX 1080Ti} & \thead{Tesla V100}\\ \hline
  \makecell[l]{Simultaneous Multiprocessors (\nsms)}  & 28 & 80 \\ \hline
  \makecell[l]{CUDA Cores per SM (\NCudaCoresPerSM)} & 128 & 64\\ \hline
  \makecell[l]{Global Memory Bandwidth\\ (\GLMemBW)} & 484 GBps & 898 GBps\\\hline
  \makecell[l]{Maximum Shared Memory Per \\Thread Block (\MaxShPerTB)} & 48 KB & 96 KB\\ \hline
  \makecell[l]{Shared Memory per SM\\ (\ShPerSM)} & \multicolumn {2}{c|}{96 KB}\\\hline
  \makecell[l]{Maximum Warps per SM \\ (\MaxWarpPerSM)} & \multicolumn {2}{c|}{64}\\\hline
  \makecell[l]{Maximum Thread Blocks per SM \\(\MaxThreadBlockPerSM)} & 16 & 32 \\\hline
  \makecell[l]{Registers per SM (\RegPerSM)} & \multicolumn {2}{c|}{65536}\\ \hline
  \makecell[l]{Maximum Registers Per Thread\\ (\MaxRegPerThread)} & \multicolumn {2}{c|}{256}\\ \hline
  \makecell[l]{Warp Size (\warpSize)} & \multicolumn {2}{c|}{32}\\ \hline
  \makecell[l]{Global Memory Transaction Size\\ (\GLMemTxSize{})} & \multicolumn {2}{c|}{\makecell[c]{32 B for L2 Cache \\ 128 B for L1 Cache}}\\
  \hline
\end{tabular}
\caption{\small Specifications of the GPUs we use in experiments.}
\label{tab:gpu-conf}
\vspace{-2em} 
\end{table}

\begin{algorithm}[thbp] 
  \small
  \caption{Cost Function} \label{algo:gpu-costfunc}
  \begin{algorithmic}[1]
      \Function{Cost} {G, tileSize, tbSize, isHybridTile, fracRegTile}
          \If {\texttt{not} constantDependenceVectors(G)} \Return $\infty$ \EndIf\label{algo:gpu-costfunc:line:align-scale}          
          \State{totalThreads $\gets$ \textsc{TotalThreads}(\textsc{GetDimSizes(G)}, tileSize)} \label{algo:config-cost:line:init-start}
          \State {warpTileSizes $\gets$ \textsc{WarpTile}(tileSizes, \textsc{WarpSizes}(tbSize))}
          \State {warpsPerTB $\gets$ \textsc{ThreadsPerTB}(tbSize) $\div$ \warpSize} \label{algo:config-cost:line:init-end}
          \State {tbPerSM $\gets$ totalThreads $\div$ \textsc{ThreadsPerTB}(tbSize) $\div$ \nsms}\label{algo:config-cost:line:tbPerSM}
          \State {warpTileVol $\gets$ \textsc{ComputeTileVol}(G, warpTileSizes)} \label{algo:config-cost:line:shared-mem-start}
          \State {totalBuff $\gets$ \textsc{NumBuffers}(G)}
          \State {shMemPerTB $\gets$ warpTileVol $\times$ warpsPerTB $\times$ totalBuff} \label{algo:config-cost:line:shared-mem-end}
          \State {shMemPerTB $\gets$ shMemPerTB $\times$(1 - fracRegTile)}
          \State {regTile$\gets$shMemPerTB$\times$fracRegTile$\div$tbSize} \label{algo:cost:line:hybrid-tile}
          \If {shMemPerTB $>$ \MaxShPerTB} \label{algo:config-cost:line:shared-mem-check}
            \Return $\infty$
          \EndIf

          \State{totalGLMemTxs $\gets$ 0}
          \ForAll {glLoad $\in$ \textsc{GetGlobalMemLoads}(G)} \label{algo:config-cost:line:glmemaccess-start}
            \State {warpLoad $\gets$ \textsc{GLLoadsInWarp}(glLoad, tileSize, tbSize)}\label{algo:config-cost:line:glwarpaccess}
            \State {glTxs $\gets$ \textsc{MinGLTxs}(warpLoad, \GLMemTxSize)}\label{algo:config-cost:line:glMemTrans}
            \State {totalGLMemTxs $\gets$ totalGLMemTxs + glTxs $\times$ tileVol} \label{algo:config-cost:line:update-total-trans}
          \EndFor \label{algo:config-cost:line:glmemaccess-end}

          \State{maxTBPerSM $\gets$ min($\frac{\ShPerSM}{\text{shMemPerTB}}$, \MaxThreadBlockPerSM)}\label{algo:config-cost:line:min-limit-maxtbpersm}
          \State{shMemOcc $\gets$ min(maxTBPerSM $\times$ warpsPerTB, \MaxWarpPerSM)}\label{algo:config-cost:line:shmemoccupancy}
          
          \State{regPerTh $\gets$ regTile + $\sum_{\text{H} \in \text{G}}$ \textsc{RegUsage}(H)} \label{algo:config-cost:line:regperthread}
          \If {regPerTh $>$ \MaxRegPerThread} \Return {$\infty$} \EndIf
          \State{maxThPerSM $\gets$ min($\frac{\RegPerSM}{\text{regPerTh}}$, \MaxThreadsPerSM)} \label{algo:config-cost:regoccupancy-start}
          \State{regOcc $\gets$ maxThPerSM $\div$ \warpSize }\label{algo:config-cost:regoccupancy-end}
          \State{occupancy $\gets$ \texttt{min}(shMemOcc, regOcc) $\div$ \MaxWarpPerSM}\label{algo:config-cost:line:occupancy}
          \State {warpBW $\gets \GLMemBW\times\warpSize{}\div\nsms\times\NCudaCoresPerSM$} \label{algo:config-cost:line:perwarpbw}  
          \State{memTime $\gets$ \GLMemTxSize$\times$totalGLMemTxs $\div$ warpBW} \label{algo:config-cost:line:memorytime}
          \State {tileVol $\gets$ \textsc{ComputeTileVol}(G, tileSizes)}
          \State{computeTime $\gets \sum_{\text{H} \in \text{G}}$ \textsc{TimePerIter}(H)$\times$tileVol}\label{algo:config-cost:line:computetimeperiter} \label{algo:config-cost:line:computetime}

          \State{shMemPerSM $\gets$ shMemPerTB $\times$ maxTBPerSM}
          \State{unallocatedShMem $\gets$ 1 $-$ shMemPerSM$\div$ \ShPerSM{}}\label{algo:config-cost:line:shmempersm}
          \State{regPerSM $\gets$ regPerTh $\times$ \MaxWarpPerSM$\times$ \warpSize}
          \State{unusedReg $\gets$ 1 $-$ regPerSM$\times$ occupancy$\div$\RegPerSM{}}\label{algo:config-cost:line:regpersm}
          \State{fracOverlap $\gets$ \textsc{OverlapComputations}(G)$\div$ tileVol}\label{algo:config-cost:line:frac-overlap}
          \State{extraTBs $\gets$ totalTB \% maxTBPerSM}\label{algo:config-cost:cleanupt-tbs}
          \State {cost = $w_1 \times $totalGLMemTxs + $w_2 \times $(1 $-$ occupancy) + $w_3 \times $ memTime $\div$ computeTime + $w_4 \times $unallocatedShMem + $w_5 \times $unusedReg + $w_6 \times $ fracOverlap + $w_7 \times $ extraTBs} \label{algo:config-cost:cost}
          \State{}\Return {cost}
      \EndFunction
        
  \end{algorithmic}
\end{algorithm}

Algorithm~\ref{algo:gpu-costfunc} is our cost function, and itt takes four
arguments: 1)~a group of stages to fuse, \textsc{G}, 2)~tile sizes, 3)~thread
block sizes, 4)~fraction of tile stored in registers, and returns the cost. The
function refers to the the hardware configuration of a GPU (
Table~\ref{tab:gpu-conf}). The
expression below calls the \textsc{Cost} function for all tile sizes, thread block
sizes, and fraction of tile stored in registers including 0.0 (hybrid tiling disabled)
and 1.0 (except the overlap in split dimension
the complete tile is stored in registers), and global memory transaction size for
both L1 and L2 global memory cache, and returns the minimum cost with the
appropriate global memory cache enabled, tile sizes, thread block sizes, and
the fraction of tile stored in registers:
\[\argmin_{\substack{\text{tileSize} \in \text{Tile Sizes},\\ \text{tbSize} \in \text{Thread Block Sizes}, \\\text{fracRegTile} \in \{0.0, 0.1, \ldots, 1.0\}, \\\texttt{GLMemTxSize} \in \{32, 128\}}} \text{\small \textsc{Cost} (G, tileSize, tbSize, fracRegTile)}\]

The \textsc{Cost} function determines the cost (line~\ref{algo:config-cost:cost}) based on
1)~the number of global memory transactions
per warp, 2)~theoretical maximum occupancy,
3)~achieved occupancy, 4)~shared memory 
usage, 5)~register usage, 6)~the fraction of redundant computations, and 
7)~the load imbalance. We calculate the weighted sum of these factors to
determine the cost.  The function also ensures the dependence vectors between all stages of a group are 
constants after alignment and scaling of 
dependencies (line~\ref{algo:gpu-costfunc:line:align-scale}). 
The function determines the dimension sizes of the group, 
total threads created,  
threads per thread block, number of warps per thread block,
and warp overlapped tile sizes (lines~\ref{algo:config-cost:line:init-start}--\ref{algo:config-cost:line:init-end}).
We distribute all thread blocks equally across all SMs (line~\ref{algo:config-cost:line:tbPerSM}). 
We retrieve the volume of each tile, the number 
of intermediate buffers, and multiply them with number of warps 
per thread block to determine shared memory usage per thread block (lines~\ref{algo:config-cost:line:shared-mem-start}--\ref{algo:config-cost:line:shared-mem-end}). 

If hybrid tiling is used, the function splits the shared memory tile into two
parts and updates the register tile (line~\ref{algo:cost:line:hybrid-tile}). We
check if the shared memory used per thread block is more than the maximum
shared memory (line~\ref{algo:config-cost:line:shared-mem-check}).

The rest of this section describes how we calculate the weight of each component
of the cost.

\paragraph{Number of Global Memory Transactions}
The cost function estimates the number of global memory transactions that
either load input images or inputs to the group (lines~\ref{algo:config-cost:line:glmemaccess-start}--\ref{algo:config-cost:line:glmemaccess-end}).
The number of global memory transactions depends on tile sizes,
thread block sizes, and the global memory transaction size. 
Higher global memory transaction size is beneficial when all values loaded from
the global memory are used by the group. If not all loaded values are used in
the group, then it is better to use a smaller transaction size. Using
\citet{pldi91:wolf}, we retrieve the loads for each global memory load for all
threads in a warp (line~\ref{algo:config-cost:line:glwarpaccess}). We coalesce
all memory loads into the minimum number of transactions
(line~\ref{algo:config-cost:line:glMemTrans}). Finally, we calculate the total
number of transactions (line~\ref{algo:config-cost:line:update-total-trans}).

\paragraph{Theoretical Occupancy} We estimate theoretical occupancy based
on shared memory and register utilization. We calculate the maximum number of
thread blocks supported by an SM based on the shared memory usage and take its
minimum with \MaxThreadBlockPerSM{}
(line~\ref{algo:config-cost:line:min-limit-maxtbpersm}). Multiplying this value
with number of warps per thread block gives the occupancy from shared memory
usage (line~\ref{algo:config-cost:line:shmemoccupancy}). We sum the register usage
of all stages in the group from in \textit{preprocessing} step to get the
register usage of the group (line~\ref{algo:config-cost:line:regperthread}). We
obtain the occupancy from register usage by determining the maximum number of
warps supported based on register usage and taking the minimum with the
\MaxWarpPerSM{}
(lines~\ref{algo:config-cost:regoccupancy-start}--\ref{algo:config-cost:regoccupancy-end}).
The ratio of minimum of both occupancies to \MaxWarpPerSM{} is the theoretical
occupancy (line~\ref{algo:config-cost:line:occupancy}).

\paragraph{Achieved Occupancy}

The cost function estimates the number of warps ready to execute at runtime as
the ratio of time spent in global memory loads to the time spent in
computations. This ratio must be decreased, since, theoretical occupancy cannot
be reached at runtime if warps spent most of their time waiting for global
memory requests to be fulfilled and an SM's compute resources are idle. To
determine the time spent in global memory loads, we divide the theoretical
global memory bandwidth equally among all SMs, and then among all warps that
can execute in parallel (line~\ref{algo:config-cost:line:perwarpbw}). Hence,
this produces the time spent in all global memory transactions
(line~\ref{algo:config-cost:line:memorytime}). We do not use a cost model to
obtain the computation time because GPU uses optimizations like pipelining
instead we obtain the execution time of each stage as mentioned in
\textit{preprocessing} step and then determine the computation time for the
group by the summing the computation time for individual stages and multiplying
that by the tile size (line~\ref{algo:config-cost:line:computetimeperiter}).

\paragraph{Shared Memory and Register Usage}

The cost function maximizes
the shared memory and register usage in addition to occupancy because while
higher occupancy can imply lower shared memory or register usage, high shared
memory or register usage can lead to lower occupancy. We calculate per thread
block shared memory usage and register usage
(line~\ref{algo:config-cost:line:shmempersm}--\ref{algo:config-cost:line:regpersm})
when all thread blocks are executing concurrently based on the occupancy.

\paragraph{Fraction of Redundant Computations} The cost function determines the
fraction of overlap (line~\ref{algo:config-cost:line:frac-overlap}).

\paragraph{Load imbalance}
The cost function minimizes the load imbalance due to when the number of thread
blocks per SMs are not always a multiple of number of thread blocks executing
concurrently per SM based on the occupancy.
Line~\ref{algo:config-cost:cleanupt-tbs} determines the extra thread blocks for each SM.

\section{Evaluation}
\label{sec:impl-eval}

In this section, we investigate the following questions:
1)~How fast is our automatic loop fusion algorithm?
2)~How does the \textit{OTPW} execution model compare to the state-of-the-art?
3)~How do \textit{OTPW with Hybrid Tiling} compare to the
state-of-art? 
4)~Why do \textit{OTPW and Hybrid Tiling} perform well?

\paragraph{Experimental Setup} We use a 3.4 GHz, quad-core Intel i5-4670 CPU
with 16GB RAM and two GPUs (each experiment uses a single GPU): an NVIDIA GTX 
1080Ti and an NVIDIA Tesla V100 (Table~\ref{tab:gpu-conf} lists their key specifications).
For our benchmarks, we use six canonical image processing applications that have
appeared in prior
work~\cite{pldi13:halide,siggraph19:halide-auto-dl,sigraph16:halide-cpu-auto,asplos15:polymage,ppopp18:polymage-dp}.
Table~\ref{tab:benchmarks} reports the number of stages and the size of the
input image for each benchmark. We compare our work to the manually-written 
schedules present in Halide repository~\cite{halide-repo}, 
Li et al.'s autoscheduler for Halide~\cite{siggraph18:halide-auto-diff}, Rawat et al.'s code generator~\cite{pact16:rawat}, and PolyMage's autotuner. We compiled Halide with LLVM
10.0.
The execution time that we report for each benchmark is the sum of execution 
time of all generated CUDA kernels (obtained using \texttt{nvprof}), and does not 
include host and device memory transfer time. We execute each benchmark for three 
samples with each sample containing 100 runs. We report the minimum of the average 
running time for each sample.

\begin{table}[t]
  \small
  \begin{tabular}{|l@{}|r|r|r|}
  \hline
  \thead{Benchmark} & \thead{Stages} & \thead{Image size (W$\times$H$\times$c)} & \thead{Fusion} \\
  \hline
  Unsharp Mask (UM) &4 & 4256$\times$2832$\times$3&0.05s\\ \hline
  Harris  Corner (HC) &11 &4256$\times$2832&0.15s\\ \hline
  Bilateral Grid (BG) &7 & 2560$\times$1536&0.02s\\ \hline
  Multiscale Interp. (MI) & 49 &2560$\times$1536$\times$3&10s\\\hline
  Camera Pipeline (CP) &32 & 2592$\times$1968&17s\\\hline
  Pyramid Blend (PB) & 44 & 3840$\times$2160$\times$3&28s\\
  \hline
\end{tabular}
\caption{\small For each benchmark, the number of stages, size of input,
and time taken for loop fusion.}
\label{tab:benchmarks}
\end{table}

\paragraph{Cost Function Weights}

The cost function that we use for automatic fusion requires several weights
that are GPU-dependent. We determine the best weights empirically using leave-one-out cross validation, since, there are small number of benchmarks. 
Table~\ref{tab:weights} shows the weights.

\begin{table}[t]
  \small
  \begin{tabular}{|l|c|c|c|c|c|c|c|}
    \hline
    & $w_1$ & $w_2$ & $w_3$ & $w_4$ & $w_5$ & $w_6$ & $w_7$\\
    \hline
    GTX 1080Ti &50 & 0.5& 45& 20& 2&100&1\\ \hline
    Tesla V100 &50 & 0.5& 60& 10& 2&100&1\\ \hline
  \end{tabular}
  \caption {\small Value of weights obtained for both GPUs.\label{tab:weights}}
  \vspace{-1em}
\end{table}

\begin{table}[t]
  \small
  \begin{tabularx}{\linewidth}{l@{}c@{\hspace{10pt}}c@{\hspace{10pt}}c@{\hspace{10pt}}c@{\hspace{10pt}}c@{\hspace{10pt}}c@{\hspace{10pt}}c@{\hspace{10pt}}}
      \toprule
      Benchmark  & \multicolumn{2}{c}{Halide} & \multicolumn{2}{c}{PolyMage-GPU} & \multicolumn{2}{c}{Speedup}\\
      \cmidrule(lr){2-3} \cmidrule(lr){4-5} \cmidrule(lr){6-7}
                          & 1080Ti & V100   & 1080Ti & V100 & 1080Ti & V100\\ \midrule
      Unsharp Mask        &1.50 &0.45 &1.00 &0.39  &1.50 &1.15\\
      Harris Corner       &1.80 &0.45 &0.80 &0.29  &2.25 &1.55\\
      Bilateral Grid      &0.40 &0.20 &0.32 &0.20  &1.25 &1.00\\
      Multi. Interp.      &1.65 &0.60 &1.26 &0.54  &1.31 &1.11\\
      Camera Pipe.        &1.90 &0.36 &1.04 &0.30  &1.83 &1.23\\
      Pyramid Blend       &5.80 &2.90 &2.90 &1.30  &2.00 &2.23\\ \midrule
      Geomean             &     &   &     &        &1.65 &1.33\\
      \bottomrule 
  \end{tabularx}
  \caption{\small Execution times (in ms) of benchmarks and speedup of PolyMage-GPU over Halide's manually written schedules on GTX 1080Ti and Tesla V100.}
  \label{tab:nvidia-1080Ti}\label{tab:nvidia-v100}
\end{table}

\pgfplotsset{width=8cm, height=4.9cm, compat=newest} 
\usepgfplotslibrary{units} 

\begin{figure}[t]
  \small
\pgfplotstableread{
X   Benchmark         1080Ti    V100
1   UM             21.1           14.8
2   HC             11.7         15.16
3   BG             1.1          2.85
4   MI             25.4           34.4
5  CP             16.2           22.23
6  PB             15.7            18.3
}\datatable
\begin{tikzpicture}
\begin{axis}[
    ylabel=label,
    xtick=data,
    xticklabels={UM, HC, BG, MI, CP, PB},
    enlarge x limits,
    ylabel style={align=left}, ylabel = Percentage of Issue Stalls \\due to \texttt{\_\_syncthreads},
    ybar=2.0pt,
    ymin=0,
    ymax=55,
    bar width=3.5pt,
    legend style={
      font=\small,
      cells={anchor=west},
      legend columns=5,
      at={(0.5,1.0)},
      anchor=north,  
      /tikz/every even column/.append style={column sep=0.2cm}
    },
    legend cell align=left,
    after end axis/.code={\draw ({rel axis cs:0,0}|-{axis cs:0,0}) -- ({rel axis cs:1,0}|-{axis cs:0,0});
        \path [anchor=base east, yshift=0.5ex]
            (rel axis cs:0,0) node [yshift=-4ex] {Benchmarks};
    },
    nodes near coords,
    every node near coord/.append style={font=\scriptsize, rotate=90, anchor=west},
    nodes near coords align={vertical},
]

\addplot [red!90!black,fill=red!90!white] table[x=X,y=1080Ti] {\datatable};
\addlegendentry{1080Ti}
\addplot [red!20!black,fill=red!20!white] table[x=X,y=V100] {\datatable};
\addlegendentry{V100}
     
\end{axis}
\end{tikzpicture}
\caption{\small Percentage of instruction issue stalls due to thread block synchronization in OTPTB (Halide) for both GTX 1080Ti and Tesla V100. OTPW execution model does not produce any synchronization based issue stalls.
    \label{fig:stalls}
    \vspace{-1em}}
\Description[]{}
\end{figure}
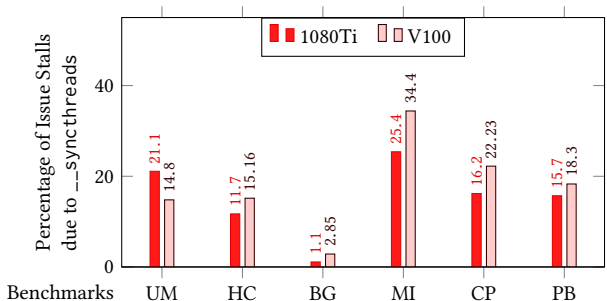

\pgfplotsset{width=17.1cm, height=4.4cm, compat=newest} 
\usepgfplotslibrary{units} 



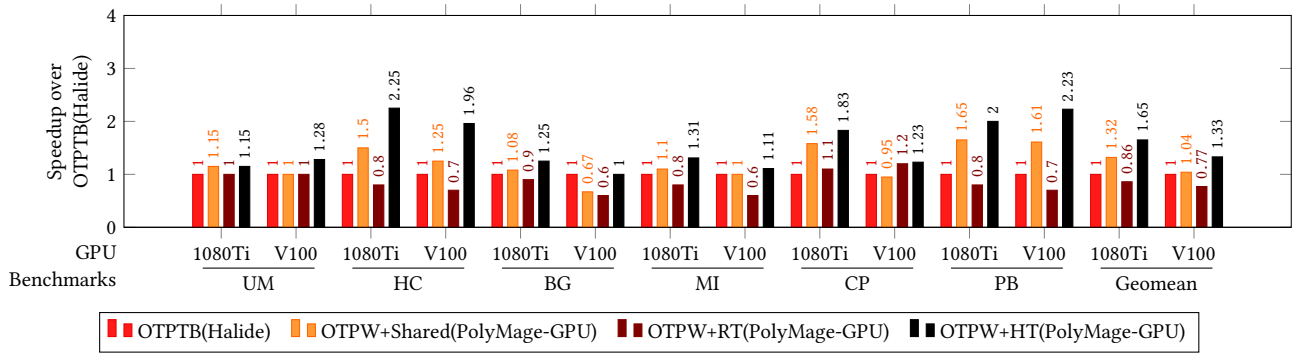
\begin{figure*}[t]
  \small
\pgfplotstableread{
X   Benchmark GPU        OTPW    HT     HM           RT
1   UM        1080Ti     1.15    1.15     1.0         1.0
2   UM        V100       1.0     1.28    1.0         1.0
3   HC        1080Ti     1.50    2.25    1.0         0.8
4   HC        V100       1.25    1.96    1.0         0.7
5   BG        1080Ti     1.08    1.25    1.0         0.9
6   BG        V100       0.67    1.00     1.0         0.6
7   MI        1080Ti     1.10    1.31    1.0         0.8
8   MI        V100       1.00    1.11    1.0         0.6
9   CP        1080Ti     1.58    1.83    1.0         1.1
10  CP        V100       0.95    1.23     1.0         1.2
11  PB        1080Ti     1.65    2.00    1.0         0.8
12  PB        V100       1.61    2.23    1.0         0.7
13  Geomean   1080Ti     1.32    1.65       1.0   0.86
14  Geomean   V100       1.04    1.33       1.0    0.77
}\datatable

\begin{tikzpicture}
\begin{axis}[
    ylabel=label,
    xtick=data,
    xticklabels={1080Ti, V100, 1080Ti, V100, 1080Ti, V100,1080Ti, V100,1080Ti, V100,1080Ti,V100, 1080Ti, V100},
    enlarge x limits,
    ylabel style={align=center}, ylabel = Speedup over\\ OTPTB(Halide),
    ybar=2.0pt,
    ymin=0,
    ymax=4,
    bar width=3.9pt,
    legend style={
      font=\small,
      cells={anchor=west},
      legend columns=5,
      at={(0.45,-0.4)},
      anchor=north,  
      /tikz/every even column/.append style={column sep=0.2cm}
    },
    legend cell align=left,
    after end axis/.code={\draw ({rel axis cs:0,0}|-{axis cs:0,0}) -- ({rel axis cs:1,0}|-{axis cs:0,0});
        \path [anchor=base east, yshift=0.5ex]
            (rel axis cs:0,0) node [yshift=-7ex] {Benchmarks}
            (rel axis cs:0,0) node [yshift=-4ex] {GPU};
    },
    nodes near coords,
    every node near coord/.append style={font=\scriptsize, rotate=90, anchor=west},
    nodes near coords align={vertical},
    draw group line={[index]1}{UM}{UM}{-4.5ex}{7pt},
    draw group line={[index]1}{HC}{HC}{-4.5ex}{7pt},
    draw group line={[index]1}{BG}{BG}{-4.5ex}{7pt},
    draw group line={[index]1}{MI}{MI}{-4.5ex}{7pt},
    draw group line={[index]1}{CP}{CP}{-4.5ex}{7pt},
    draw group line={[index]1}{PB}{PB}{-4.5ex}{7pt},
    draw group line={[index]1}{Geomean}{Geomean}{-4.5ex}{7pt}
]

\addplot [red!90!black,fill=red!90!white] table[x=X,y=HM] {\datatable};
\addlegendentry{OTPTB(Halide)}
\addplot [red!20!orange,fill=orange!80!white]table[x=X,y=OTPW] {\datatable};
\addlegendentry{OTPW+Shared(PolyMage-GPU)}
\addplot [red!50!black,fill=red!50!black] table[x=X,y=RT] {\datatable};
\addlegendentry{OTPW+RT(PolyMage-GPU)}
\addplot [black!80!black,fill=black!80!black] table[x=X,y=HT] {\datatable};
\addlegendentry{OTPW+HT(PolyMage-GPU)}
     
\end{axis}
\end{tikzpicture}
\caption{\small Times relative to OTPTB(Halide) on GTX 1080Ti and Tesla V100. 
OTPTB(Halide) are the manually written schedules in Halide following \textit{OTPTB} execution model. OTPW+Shared(PolyMage-GPU) is the implementation of \textit{OTPW} execution 
model in PolyMage-GPU with tiles stored only in shared memory. OTPW+RT(PolyMage-GPU) is the implementation of \textit{OTPW} with tiles stored only in registers in PolyMage-GPU. OTPW+HT(PolyMage-GPU) is the implementation of \textit{OTPW} with Hybrid Tiling in PolyMage-GPU. 
\vspace{-0.5em}
    \label{fig:all-results}}
\Description[]{}
\end{figure*}

\subsection{Automatic Fusion Time}

We first measure the time it takes for automatic fusion to process each
benchmark program to find an optimal schedule. We use Bounded DP Fusion~\cite{ppopp18:polymage-dp}. 
to search for (i)~thread block sizes (as a multiple of \warpSize{}), and (ii) tile
sizes from 1 to 32 in each dimension. The \emph{Fusion} column in
Table~\ref{tab:benchmarks} shows the time taken, which ranges from less than a
second to up to 30 seconds for benchmarks with a few dozen stages. In contrast, the
 PolyMage autotuner can take up to 20 hours (Section~\ref{sec:other-comparison}). Thus,
our approach to automatic fusion is significantly faster.

\subsection{Performance Evaluation}

We now evaluate the performance of OTPW with hybrid tiling and the loop fusion
algorithm, which we implement in a tool that we call \textbf{PolyMage-GPU}.\footnote{The generated CUDA
10.0 is compiled using \texttt{nvcc -O3 -arch=compute\_61 -code=sm\_61} on the GTX 1080Ti and \texttt{nvcc -O3
-arch=compute\_70 -code=sm\_70} on the Tesla V100.}. 
We compare our work to the manually-written 
schedules present in the Halide repository~\cite{halide-repo}. However, we wrote the
schedule for \textit{Pyramid Blend} ourselves, since it was not available.





  

Table~\ref{tab:nvidia-1080Ti} shows the absolute execution times of
PolyMage-GPU and Halide and the speedup of PolyMage-GPU over Halide on both
GPUs. On every benchmark, PolyMage-GPU is at least as fast as Halide, and in
many cases, significantly faster. PolyMage-GPU is faster than manually written
schedules in Halide with a geomean speedup of 1.65$\times$ and 1.33$\times$ on
the GTX 1080Ti and Tesla V100 respectively. In general, PolyMage-GPU outperforms Halide
because its fusion algorithm chooses better thread block and tile sizes, and the
runtime technique has lower synchronization cost, decreased shared memory
usage, and improved occupancy. The only exception is the \textit{Bilateral
Grid} benchmark on V100, where Halide's manual schedules are competitive with
PolyMage-GPU because Halide can fuse the histogram stage, which performs a reduction, with subsequent blurring stages~\cite{halide-reductions}, whereas PolyMage-GPU cannot.

\begin{table}[t]
  \small
  \centering
  \begin{tabularx}{\linewidth}{@{}l@{}c@{}c@{}c@{}c@{}c@{}c@{}c}
      \toprule
      Benchmark           &\multicolumn{2}{@{}c}{Decrease in} & \multicolumn{2}{@{}c}{Increase in} & \multicolumn{2}{c}{Reasons}\\
                          & \multicolumn{2}{@{}c}{Global Loads (\%)} & \multicolumn{2}{@{}c}{Occupancy (\%)} \\
      \cmidrule(lr){2-3} \cmidrule(lr){4-5} \cmidrule(lr){6-7}
      & 1080Ti & V100   & 1080Ti & V100 & 1080Ti & V100\\\hline
      Unsharp Mask        & 2.51& 3.10      & 0.00 & 0.00  &$\downarrow$L&$\downarrow$L\\
      Harris Corner       & 20.0 & 31.2     & 9.10 & 0.00 &$\downarrow$L+$\uparrow$O&$\downarrow$L\\
      Bilateral Grid      & 4.50 & 3.60     & 0.00 & 0.00  &$\downarrow$L&$\downarrow$L&\\
      Multiscale Interp.  & 5.30 & 13.20 & 0.00 & 10.0 &$\downarrow$L&$\downarrow$+$\uparrow$O\\
      Camera Pipeline     & 5.21 & 0.00 & 1.70 & 16.6 &$\downarrow$L+$\uparrow$O&$\uparrow$O\\
      Pyramid Blend       & 9.12 & 7.40 & -5.40 & 13.8 &$\downarrow$L&$\downarrow$L+$\uparrow$O\\
      \bottomrule 
  \end{tabularx}
  \caption{\small Decrease in the number of global memory loads (in \%) and increase in achieved occupancy (in \%) of code generated using OTPW and Hybrid Tiling over code generated using OTPW on GTX 1080Ti and Tesla V100. Last columns lists the reasons for the increase in performance on both GPUs. $\downarrow$L represents decrease in number of global memory loads and $\uparrow$O represents increase in the achieved occupancy.
  \vspace{-1.5em}
  }
  \label{tab:reasons}
\end{table}

\subsubsection{Performance Analysis}

To study why the OTPW model outperforms the OTPTB, we first investigate
instruction stalls due to thread block synchronization.
Figure~\ref{fig:stalls} shows that on most benchmarks, a significant fraction
of GPU instructions stall due to thread block synchronization. These stalls
lead to idle resources which slows down the computation. In contrast, the OTPW
model does not employ thread block synchronization at all.

Next, we investigate the impact of
hybrid tiling. To do so, we modify PolyMage-GPU to disable hybrid tiling: it
still uses the OTPW model, but store tiles either entirely in shared memory
(\textbf{OTPW+Shared}) or entirely in registers (\textbf{OTPW+RT}).
Figure~\ref{fig:all-results} compares the performance of hybrid tiling
(\textbf{OTPW+HT}), with the two aforementioned approaches, using thread block
tiling (\textbf{OTPTB}) as the baseline. On the GTX 1080Ti, \textit{OTPW+Shared}
provides a geomean speedup of 1.32$\times$ over \textit{OTPTB}: it has no
instruction issue stalls, and better grouping with thread block sizes and tile
sizes. On the Tesla V100, all benchmarks perform at least as well as
\textit{OTPTB} (geomean speedup of 1.04$\times$), with the exception of
\textit{Bilateral Grid}. On Bilateral Grid, Halide's manual schedule fuses the
reduction stage with the next blurring stage, but the PolyMage compiler cannot.
On the V100, \textit{OTPW+Shared} gives the same performance as Halide for
\textit{Camera Pipeline} because the manually written schedule performs
significant inlining, which the PolyMage compiler cannot do.

Overall, \textit{OTPW+HT} improves the performance of \textit{OTPW+Shared}, with geomean speedups of 1.25$\times$ (GTX 1080Ti) and
1.28$\times$ (Tesla V100). 
To investigate further, Table~\ref{tab:reasons} reports how Hybrid Tiling
decreases the number of global memory loads, and increases achieved occupancy
in contrast to \textit{OTPW+Shared}.

On both GPUs, Hybrid Tiling improves the performance of \textit{Unsharp
Mask} and \textit{Harris Corner} by decreasing the number of global memory reads, since
hybrid tiling allows larger tile sizes, thereby decreasing the number of
overlapping computations. Moreover, Hybrid Tiling increases the occupancy in
\textit{Harris Corner} by decreasing shared memory usage. For example, on the GTX
1080Ti, \textit{OTPW+Shared} limits the tile size in \textit{Harris Corner} to
4$\times$1. However, Hybrid Tiling allows 10$\times$1 tiles, with 
equally divided among shared memory and registers.
On both GPUs, the performance of \textit{Bilateral Grid} also improves due to
increased tile sizes, and thus fewer overlapping computations, and fewer
global memory loads. For the other three benchmarks, the performance
improvement is either due to increase in tile sizes, improved occupancy, or
both. On the Tesla V100 \textit{Multiscale Interp.} performs better due to
 a decrease in global memory loads, and an increase in achieved occupancy. The best
performing tile sizes of \textit{Multiscale Interp.} decreases the number of
overlapping computations but requires more shared memory than the per
thread block shared memory limit of Tesla V100, which is decreased to half with
Hybrid Tiling. Similarly, on the GTX 1080Ti, opportunity for larger tile size in
\textit{Multiscale Interp.} due to Hybrid Tiling decreased the number of
overlapping computations. On GTX 1080Ti, Hybrid Tiling decreases the global
memory loads and slightly increases the occupancy in \textit{Camera Pipeline}.
On Tesla V100, tile sizes of \textit{Pyramid Blend} were increased in Hybrid
Tiling due to extra storage for registers available, hence, leading to low
overlapping computation, thereby, less global memory loads and increased
occupancy. In summary, Hybrid Tiling provides performance improvements due to
two major reasons: 1)~the extra storage afforded by registers allows larger tiles, which
decreases the number of overlapping computations, which in turn, decreases the
number of global memory loads, and 2)~storing portions of tiles registers
decreases the allocated shared memory, hence increases the theoretical and
achieved occupancy.

Finally, we note that \textit{OTPW+HT} and \textit{OTPW+Shared} are both faster \textit{OTPW+RT}. Register-only tiles
forces PolyMage-GPU to use tiny tiles, which results in a lot of redundant
computations.

\subsubsection{Comparison with other techniques}
\label{sec:other-comparison}


\paragraph{Rawat et al.'s Code Generator}

We compare PolyMage-GPU to the code generator of Rawat et
al.~\cite{pact16:rawat}. PolyMage-GPU provides a geomean speedup of 1.6$\times$
and 1.7$\times$ on the GTX 1080Ti and Tesla V100 respectively. Rawat et al.'s
technique has three major drawbacks. First, in their execution model each
thread processes exactly one point, whereas PolyMage-GPU does not have this
limitation. Thus PolyMage-GPU supports larger tile sizes, and is able to use
Hybrid Tiling. Second, since their sliding window technique streams overlapped
tiles in one dimension, there is no parallelism in that dimension, thereby
leading to significant decrease in total parallelism. Finally, unlike
PolyMage-GPU, their cost function is geared towards minimizing the data
movement with optimizing shared memory and register usage, hence, does not
consider thr number of global memory transactions and achieved occupancy.
Hence, our approach decreases the amount of overlapping computations without
decreasing in parallelism.

\paragraph{Halide's Gradient GPU autoscheduler}

We compare PolyMage-GPU wih Halide's Gradient GPU
autoscheduler~\cite{siggraph18:halide-auto-diff}. To use
Halide's latest code generation features, we used the schedules generated by the
autoscheduler in the latest Halide version. We found that PolyMage-GPU provides a geomean
speedup of 2.42$\times$ and 2.35$\times$ on the GTX 1080Ti and Tesla V100
respectively. We believe this difference occurs because the thread block sizes
picked by PolyMage-GPU are better suited for both GPUs than the hard coded
thread block sizes used by Halide's Gradient GPU autoscheduler.

\paragraph{PolyMage's Autotuner}

We also compare to PolyMage's image processing
autotuner~\cite{asplos15:polymage}. We added support for OTPW and Hybrid Tiling
in the autotuner. The model based autotuner takes tile sizes, thread block
sizes, and an overlap threshold. To reduce the search space, PolyMage assigns
same tile sizes to all groups and using a greedy approach selects stages to
fuse. The greedy approach groups all stages till the fraction of overlap is
within a given threshold. Similar to ~\cite{asplos15:polymage}, we use same
overlap threshold values: 0.2, 0.4, and 0.5. We use tile sizes from 1 to 32 in
each dimension, and thread block size of 1 to 512 in each dimension.
PolyMage-GPU is 4.5$\times$ and 3.3$\times$ faster than PolyMage-A on GTX
1080Ti and Tesla V100. PolyMage-A runs till 20 hours to generate these
schedules, while PolyMage-GPU runs in seconds. Since, PolyMage-A decreases the
search space by selecting the same tile size and thread block sizes for all
groups, all schedules are not explored. Hence, PolyMage-A does not find the
same schedules as PolyMage-GPU.

\section{Related Work}
\label{sec:related-work}

State-of-the-art DSLs for image processing programs all employ loop fusion and
overlapped tiling to increase locality between
stages~\cite{pldi13:halide,asplos15:polymage,gpgpu8:forma}. Halide and Forma
use GPUs and execute one overlapped tile per thread block. Halide's original
CPU autoscheduler~\cite{sigraph16:halide-cpu-auto} uses a greedy algorithm,
whereas Dynamic Programming Fusion~\cite{ppopp18:polymage-dp} efficiently
enumerates all possible fusion choices for a CPU. Halide has a newer
autoscheduler~\cite{siggraph19:halide-auto-dl} that uses beam search with a
learned cost model for CPUs. Halide's Gradient GPU autoscheduler~\cite{siggraph18:halide-auto-diff}
is a GPU autoscheduler for Halide that performs greedy function inlining
and loop fusion with hard-coded thread block sizes for each tile. In
Section~\ref{sec:other-comparison}, we compare our work to some of these
autoscheduler.

Versapipe~\cite{Versapipe} exploits both task and data parallelism on GPUs
by assigning tasks to persistent threads based on their SM ID. HiWayLib~\cite{HiWayLib} presents a way to efficiently run
pipelined computations that require significant communication between CPUs and
GPUs. In contrast, our work focuses on improving the performance of image
processing pipelines, which are data parallel applications, and we require
all data to fit on the GPU. We employ a warp-centric approach and use a
cost function to select stages for fusion. The aforementioned approaches
would complement our work.

Several
techniques
support the parallel execution of stencil computations on GPUs, using the \textit{Overlap
tile per thread block (OTPTB)} model~\cite{pact16:rawat,ics12:holewinski,ipdps19:rawat,ipdps19:artemis,taco2019:flextended-tiles}.
Rawat et al.~\cite{pact16:rawat} use a sliding window on one spatial dimension and
overlap tiling on the others to eliminate some redundant computations in
Overtile~\cite{ics12:holewinski}. \textit{Hybrid hexagonal classic tiling}
~\cite{cgo14:hht} also executes one tile per thread block. Flextended
Tiles~\cite{taco2019:flextended-tiles} uses rectangle trapezoid tiling to obtain
tighter overlapped tile bounds. Artemis~\cite{ipdps19:artemis} is a DSL that allows
an expert to guide challenging code optimizations using bottleneck analysis and
tunable code parameters. Artemis and Flextended tiles are complementary to our
work. These approach supports expression inlining, which pass the value of producer to consumer through a register within the same thread.
However, none of these employ the \emph{overlapped tile per warp (OTPW)}
model and \emph{hybrid tiling}, which stores portions of tiles in registers that is shared among threads of a warp.

In 2009, \citet{isca09:hong} presented a general analytical model to predict
the performance of GPU kernels. However, recent advances in GPU architectures,
including changes to their memory hierarchy, have made their model out of date.
\citet{ppopp17:prajapati} present an analytical model for predicting the
runtime of stencil computations on GPUs (tiled using \cite{cgo14:hht}). That
model considers shared memory usage, theoretical occupancy, and warp switching.
However, it omits several key factors, including register usage, the number of
global memory transactions, achieved occupancy, and thread block sizes, which
our model considers.


Halide exposes warp shuffle instructions, which makes it possible to store
portions of a tile in registers~\cite{halide-store-in}. However, Halide
restricts the size of the innermost dimension to be less than warp size, and
cannot store tiles in both registers and shared memory. Other systems employ
in-register storage and warp shuffles to improve the performance of GPU
kernels~\cite{ics16:fast-binary-fields,ics18:warp-consolidation,ipdps17:wang,cgo2019:gonzalo,ics19:polymage-life,asplos19:swizzle-inventor}.
Our work allows multiple warps per thread block, allows the innermost dimension
to have an arbitrary size, and is a hybrid technique that stores tiles in both
registers and shared memory. To the best of our knowledge, this combination has
not been presented in prior work.



\section{Conclusion}
\label{sec:conclusion}

This paper presents 1)~an execution model for image processing pipelines on
GPUs that executes one overlapped tile per warp, 2)~\emph{hybrid tiling}, which
allows portions of overlapped tiles to be stored in either registers or shared
memory, and 3)~ an automatic loop fusion technique for GPUs that considers
several key factors that affect the performance of GPU kernels. These
techniques use low cost synchronization, improves occupancy, and allows larger
tiles that require fewer overlapping computations. We implement these
techniques in PolyMage-GPU, which is a new GPU backend for the PolyMage DSL.
Using several benchmarks, we show that our work achieves significant speedups
over manually-written schedules.

\section*{Acknowledgements}

This work was partially supported by the National Science Foundation under grant CCF-1717636.

\bibliographystyle{ACM-Reference-Format}
\bibliography{papers}


\begin{thebibliography}{35}


\ifx \showCODEN    \undefined \def \showCODEN     #1{\unskip}     \fi
\ifx \showDOI      \undefined \def \showDOI       #1{#1}\fi
\ifx \showISBNx    \undefined \def \showISBNx     #1{\unskip}     \fi
\ifx \showISBNxiii \undefined \def \showISBNxiii  #1{\unskip}     \fi
\ifx \showISSN     \undefined \def \showISSN      #1{\unskip}     \fi
\ifx \showLCCN     \undefined \def \showLCCN      #1{\unskip}     \fi
\ifx \shownote     \undefined \def \shownote      #1{#1}          \fi
\ifx \showarticletitle \undefined \def \showarticletitle #1{#1}   \fi
\ifx \showURL      \undefined \def \showURL       {\relax}        \fi
\providecommand\bibfield[2]{#2}
\providecommand\bibinfo[2]{#2}
\providecommand\natexlab[1]{#1}
\providecommand\showeprint[2][]{arXiv:#2}

\bibitem[\protect\citeauthoryear{??}{cud}{[n. d.]a}]%
        {cuda-guide}
 \bibinfo{year}{[n. d.]}\natexlab{a}.
\newblock \bibinfo{title}{CUDA C Programming Guide}.
\newblock
\newblock
\urldef\tempurl%
\url{https://docs.nvidia.com/cuda/cuda-c-programming-guide/}
\showURL{%
\tempurl}


\bibitem[\protect\citeauthoryear{??}{hal}{[n. d.]a}]%
        {halide-repo}
 \bibinfo{year}{[n. d.]}\natexlab{a}.
\newblock \bibinfo{title}{Halide}.
\newblock
\newblock
\newblock
\shownote{\url{https://github.com/halide/Halide/} \\ {\footnotesize commit
  52da814a2c3c4af78125757385a8a86efdde3234}.}


\bibitem[\protect\citeauthoryear{??}{hal}{[n. d.]b}]%
        {halide-store-in}
 \bibinfo{year}{[n. d.]}\natexlab{b}.
\newblock \bibinfo{title}{Halide}.
\newblock
\newblock
\newblock
\shownote{\url{https://github.com/halide/Halide/} \\ {\footnotesize commit
  59bca3c8e535f7f99c90efd1d932db934f9c01b6}.}


\bibitem[\protect\citeauthoryear{??}{cud}{[n. d.]b}]%
        {cuda-warp-level-primitives}
 \bibinfo{year}{[n. d.]}\natexlab{b}.
\newblock \bibinfo{title}{Using CUDA Warp-Level Primitives}.
\newblock
\newblock
\newblock
\shownote{\url{https://devblogs.nvidia.com/using-cuda-warp-level-primitives/}.}


\bibitem[\protect\citeauthoryear{??}{amd}{[n. d.]}]%
        {amd-hip-warp-shuffle}
 \bibinfo{year}{[n. d.]}\natexlab{}.
\newblock \bibinfo{title}{Warp Shuffle Functions in AMD HIP}.
\newblock
\newblock
\urldef\tempurl%
\url{https://github.com/ROCm-Developer-Tools/HIP/blob/master/docs/markdown/hip_kernel_language.md#warp-shuffle-functions}
\showURL{%
\tempurl}


\bibitem[\protect\citeauthoryear{Adams, Ma, Anderson, Baghdadi, Li, Gharbi,
  Steiner, Johnson, Fatahalian, Durand, and Ragan-Kelley}{Adams
  et~al\mbox{.}}{2019}]%
        {siggraph19:halide-auto-dl}
\bibfield{author}{\bibinfo{person}{Andrew Adams}, \bibinfo{person}{Karima Ma},
  \bibinfo{person}{Luke Anderson}, \bibinfo{person}{Riyadh Baghdadi},
  \bibinfo{person}{Tzu-Mao Li}, \bibinfo{person}{Michael Gharbi},
  \bibinfo{person}{Benoit Steiner}, \bibinfo{person}{Steven Johnson},
  \bibinfo{person}{Kayvon Fatahalian}, \bibinfo{person}{Fredo Durand}, {and}
  \bibinfo{person}{Jonathan Ragan-Kelley}.} \bibinfo{year}{2019}\natexlab{}.
\newblock \showarticletitle{Learning to Optimize Halide with Tree Search and
  Random Programs}.
\newblock \bibinfo{journal}{\emph{ACM Trans. Graph.}} (\bibinfo{year}{2019}).
\newblock


\bibitem[\protect\citeauthoryear{Aggarwal and Bondhugula}{Aggarwal and
  Bondhugula}{2019}]%
        {ics19:polymage-life}
\bibfield{author}{\bibinfo{person}{Karan Aggarwal} {and} \bibinfo{person}{Uday
  Bondhugula}.} \bibinfo{year}{2019}\natexlab{}.
\newblock \showarticletitle{Optimizing the Linear Fascicle Evaluation Algorithm
  for Many-core Systems}. In \bibinfo{booktitle}{\emph{Proceedings of the ACM
  International Conference on Supercomputing}} \emph{(\bibinfo{series}{ICS
  '19})}.
\newblock


\bibitem[\protect\citeauthoryear{Baskaran, Bondhugula, Krishnamoorthy,
  Ramanujam, Rountev, and Sadayappan}{Baskaran et~al\mbox{.}}{2008}]%
        {ics08:baskaran}
\bibfield{author}{\bibinfo{person}{Muthu~Manikandan Baskaran},
  \bibinfo{person}{Uday Bondhugula}, \bibinfo{person}{Sriram Krishnamoorthy},
  \bibinfo{person}{J. Ramanujam}, \bibinfo{person}{Atanas Rountev}, {and}
  \bibinfo{person}{P. Sadayappan}.} \bibinfo{year}{2008}\natexlab{}.
\newblock \showarticletitle{A Compiler Framework for Optimization of Affine
  Loop Nests for Gpgpus}. In \bibinfo{booktitle}{\emph{Proceedings of the 22nd
  Annual International Conference on Supercomputing}}
  \emph{(\bibinfo{series}{ICS '08})}.
\newblock


\bibitem[\protect\citeauthoryear{Ben-Sasson, Hamilis, Silberstein, and
  Tromer}{Ben-Sasson et~al\mbox{.}}{2016}]%
        {ics16:fast-binary-fields}
\bibfield{author}{\bibinfo{person}{Eli Ben-Sasson}, \bibinfo{person}{Matan
  Hamilis}, \bibinfo{person}{Mark Silberstein}, {and} \bibinfo{person}{Eran
  Tromer}.} \bibinfo{year}{2016}\natexlab{}.
\newblock \showarticletitle{Fast Multiplication in Binary Fields on GPUs via
  Register Cache}. In \bibinfo{booktitle}{\emph{Proceedings of the 2016
  International Conference on Supercomputing}} \emph{(\bibinfo{series}{ICS
  '16})}.
\newblock


\bibitem[\protect\citeauthoryear{De~Gonzalo, Huang, G\'{o}mez-Luna, Hammond,
  Mutlu, and Hwu}{De~Gonzalo et~al\mbox{.}}{2019}]%
        {cgo2019:gonzalo}
\bibfield{author}{\bibinfo{person}{Simon~Garcia De~Gonzalo},
  \bibinfo{person}{Sitao Huang}, \bibinfo{person}{Juan G\'{o}mez-Luna},
  \bibinfo{person}{Simon Hammond}, \bibinfo{person}{Onur Mutlu}, {and}
  \bibinfo{person}{Wen-mei Hwu}.} \bibinfo{year}{2019}\natexlab{}.
\newblock \showarticletitle{Automatic Generation of Warp-level Primitives and
  Atomic Instructions for Fast and Portable Parallel Reduction on GPUs}. In
  \bibinfo{booktitle}{\emph{Proceedings of the 2019 IEEE/ACM International
  Symposium on Code Generation and Optimization}} \emph{(\bibinfo{series}{CGO
  2019})}.
\newblock


\bibitem[\protect\citeauthoryear{Grosser, Cohen, Holewinski, Sadayappan, and
  Verdoolaege}{Grosser et~al\mbox{.}}{2014}]%
        {cgo14:hht}
\bibfield{author}{\bibinfo{person}{Tobias Grosser}, \bibinfo{person}{Albert
  Cohen}, \bibinfo{person}{Justin Holewinski}, \bibinfo{person}{P. Sadayappan},
  {and} \bibinfo{person}{Sven Verdoolaege}.} \bibinfo{year}{2014}\natexlab{}.
\newblock \showarticletitle{Hybrid Hexagonal/Classical Tiling for GPUs}. In
  \bibinfo{booktitle}{\emph{Proceedings of Annual IEEE/ACM International
  Symposium on Code Generation and Optimization}} \emph{(\bibinfo{series}{CGO
  '14})}.
\newblock


\bibitem[\protect\citeauthoryear{Holewinski, Pouchet, and
  Sadayappan}{Holewinski et~al\mbox{.}}{2012}]%
        {ics12:holewinski}
\bibfield{author}{\bibinfo{person}{Justin Holewinski},
  \bibinfo{person}{Louis-No\"{e}l Pouchet}, {and} \bibinfo{person}{P.
  Sadayappan}.} \bibinfo{year}{2012}\natexlab{}.
\newblock \showarticletitle{High-performance Code Generation for Stencil
  Computations on GPU Architectures}. In \bibinfo{booktitle}{\emph{Proceedings
  of the 26th ACM International Conference on Supercomputing}}
  \emph{(\bibinfo{series}{ICS '12})}.
\newblock


\bibitem[\protect\citeauthoryear{Hong and Kim}{Hong and Kim}{2009}]%
        {isca09:hong}
\bibfield{author}{\bibinfo{person}{Sunpyo Hong} {and} \bibinfo{person}{Hyesoon
  Kim}.} \bibinfo{year}{2009}\natexlab{}.
\newblock \showarticletitle{An Analytical Model for a GPU Architecture with
  Memory-level and Thread-level Parallelism Awareness}. In
  \bibinfo{booktitle}{\emph{Proceedings of the 36th Annual International
  Symposium on Computer Architecture}} \emph{(\bibinfo{series}{ISCA '09})}.
\newblock


\bibitem[\protect\citeauthoryear{Jangda and Bondhugula}{Jangda and
  Bondhugula}{2018}]%
        {ppopp18:polymage-dp}
\bibfield{author}{\bibinfo{person}{Abhinav Jangda} {and} \bibinfo{person}{Uday
  Bondhugula}.} \bibinfo{year}{2018}\natexlab{}.
\newblock \showarticletitle{An Effective Fusion and Tile Size Model for
  Optimizing Image Processing Pipelines}. In
  \bibinfo{booktitle}{\emph{Proceedings of the 23rd ACM SIGPLAN Symposium on
  Principles and Practice of Parallel Programming}}
  \emph{(\bibinfo{series}{PPoPP '18})}.
\newblock


\bibitem[\protect\citeauthoryear{Jangda and Guha}{Jangda and Guha}{2020}]%
        {jangda2020modelbased}
\bibfield{author}{\bibinfo{person}{Abhinav Jangda} {and} \bibinfo{person}{Arjun
  Guha}.} \bibinfo{year}{2020}\natexlab{}.
\newblock \bibinfo{title}{Model-Based Warp Overlapped Level Tiling for Image
  Processing Programs on GPUs}.
\newblock
\newblock
\showeprint[arxiv]{cs.PL/1909.07190}


\bibitem[\protect\citeauthoryear{Li, Liu, Wang, Barker, and Song}{Li
  et~al\mbox{.}}{2018b}]%
        {ics18:warp-consolidation}
\bibfield{author}{\bibinfo{person}{Ang Li}, \bibinfo{person}{Weifeng Liu},
  \bibinfo{person}{Linnan Wang}, \bibinfo{person}{Kevin Barker}, {and}
  \bibinfo{person}{Shuaiwen~Leon Song}.} \bibinfo{year}{2018}\natexlab{b}.
\newblock \showarticletitle{Warp-Consolidation: A Novel Execution Model for
  GPUs}. In \bibinfo{booktitle}{\emph{Proceedings of the 2018 International
  Conference on Supercomputing}} \emph{(\bibinfo{series}{ICS '18})}.
\newblock


\bibitem[\protect\citeauthoryear{Li, Gharbi, Adams, Durand, and
  Ragan-Kelley}{Li et~al\mbox{.}}{2018a}]%
        {siggraph18:halide-auto-diff}
\bibfield{author}{\bibinfo{person}{Tzu-Mao Li}, \bibinfo{person}{Micha\"{e}l
  Gharbi}, \bibinfo{person}{Andrew Adams}, \bibinfo{person}{Fr{\'e}do Durand},
  {and} \bibinfo{person}{Jonathan Ragan-Kelley}.}
  \bibinfo{year}{2018}\natexlab{a}.
\newblock \showarticletitle{Differentiable Programming for Image Processing and
  Deep Learning in Halide}.
\newblock \bibinfo{journal}{\emph{ACM Trans. Graph.}} (\bibinfo{year}{2018}).
\newblock


\bibitem[\protect\citeauthoryear{Mullapudi, Adams, Sharlet, Ragan-Kelley, and
  Fatahalian}{Mullapudi et~al\mbox{.}}{2016}]%
        {sigraph16:halide-cpu-auto}
\bibfield{author}{\bibinfo{person}{Ravi~Teja Mullapudi},
  \bibinfo{person}{Andrew Adams}, \bibinfo{person}{Dillon Sharlet},
  \bibinfo{person}{Jonathan Ragan-Kelley}, {and} \bibinfo{person}{Kayvon
  Fatahalian}.} \bibinfo{year}{2016}\natexlab{}.
\newblock \showarticletitle{Automatically Scheduling Halide Image Processing
  Pipelines}.
\newblock \bibinfo{journal}{\emph{ACM Trans. Graph.}} (\bibinfo{year}{2016}).
\newblock


\bibitem[\protect\citeauthoryear{Mullapudi, Vasista, and Bondhugula}{Mullapudi
  et~al\mbox{.}}{2015}]%
        {asplos15:polymage}
\bibfield{author}{\bibinfo{person}{Ravi~Teja Mullapudi}, \bibinfo{person}{Vinay
  Vasista}, {and} \bibinfo{person}{Uday Bondhugula}.}
  \bibinfo{year}{2015}\natexlab{}.
\newblock \showarticletitle{PolyMage: Automatic Optimization for Image
  Processing Pipelines}. In \bibinfo{booktitle}{\emph{Proceedings of the
  Twentieth International Conference on Architectural Support for Programming
  Languages and Operating Systems}} \emph{(\bibinfo{series}{ASPLOS '15})}.
\newblock


\bibitem[\protect\citeauthoryear{Phothilimthana, Elliott, Wang, Jangda,
  Hagedorn, Barthels, Kaufman, Grover, Torlak, and Bodik}{Phothilimthana
  et~al\mbox{.}}{2019}]%
        {asplos19:swizzle-inventor}
\bibfield{author}{\bibinfo{person}{Phitchaya~Mangpo Phothilimthana},
  \bibinfo{person}{Archibald~Samuel Elliott}, \bibinfo{person}{An Wang},
  \bibinfo{person}{Abhinav Jangda}, \bibinfo{person}{Bastian Hagedorn},
  \bibinfo{person}{Henrik Barthels}, \bibinfo{person}{Samuel~J. Kaufman},
  \bibinfo{person}{Vinod Grover}, \bibinfo{person}{Emina Torlak}, {and}
  \bibinfo{person}{Rastislav Bodik}.} \bibinfo{year}{2019}\natexlab{}.
\newblock \showarticletitle{Swizzle Inventor: Data Movement Synthesis for GPU
  Kernels}. In \bibinfo{booktitle}{\emph{Proceedings of the Twenty-Fourth
  International Conference on Architectural Support for Programming Languages
  and Operating Systems}} \emph{(\bibinfo{series}{ASPLOS '19})}.
\newblock


\bibitem[\protect\citeauthoryear{Prajapati, Ranasinghe, Rajopadhye, Andonov,
  Djidjev, and Grosser}{Prajapati et~al\mbox{.}}{2017}]%
        {ppopp17:prajapati}
\bibfield{author}{\bibinfo{person}{Nirmal Prajapati}, \bibinfo{person}{Waruna
  Ranasinghe}, \bibinfo{person}{Sanjay Rajopadhye}, \bibinfo{person}{Rumen
  Andonov}, \bibinfo{person}{Hristo Djidjev}, {and} \bibinfo{person}{Tobias
  Grosser}.} \bibinfo{year}{2017}\natexlab{}.
\newblock \showarticletitle{Simple, Accurate, Analytical Time Modeling and
  Optimal Tile Size Selection for GPGPU Stencils}. In
  \bibinfo{booktitle}{\emph{Proceedings of the 22nd ACM SIGPLAN Symposium on
  Principles and Practice of Parallel Programming}}
  \emph{(\bibinfo{series}{PPoPP '17})}.
\newblock


\bibitem[\protect\citeauthoryear{Ragan-Kelley, Barnes, Adams, Paris, Durand,
  and Amarasinghe}{Ragan-Kelley et~al\mbox{.}}{2013}]%
        {pldi13:halide}
\bibfield{author}{\bibinfo{person}{Jonathan Ragan-Kelley},
  \bibinfo{person}{Connelly Barnes}, \bibinfo{person}{Andrew Adams},
  \bibinfo{person}{Sylvain Paris}, \bibinfo{person}{Fr{\'e}do Durand}, {and}
  \bibinfo{person}{Saman Amarasinghe}.} \bibinfo{year}{2013}\natexlab{}.
\newblock \showarticletitle{Halide: A Language and Compiler for Optimizing
  Parallelism, Locality, and Recomputation in Image Processing Pipelines}. In
  \bibinfo{booktitle}{\emph{Proceedings of the 34th ACM SIGPLAN Conference on
  Programming Language Design and Implementation}} \emph{(\bibinfo{series}{PLDI
  '13})}.
\newblock


\bibitem[\protect\citeauthoryear{Ravishankar, Holewinski, and
  Grover}{Ravishankar et~al\mbox{.}}{2015}]%
        {gpgpu8:forma}
\bibfield{author}{\bibinfo{person}{Mahesh Ravishankar}, \bibinfo{person}{Justin
  Holewinski}, {and} \bibinfo{person}{Vinod Grover}.}
  \bibinfo{year}{2015}\natexlab{}.
\newblock \showarticletitle{Forma: A DSL for Image Processing Applications to
  Target GPUs and Multi-core CPUs}. In \bibinfo{booktitle}{\emph{Proceedings of
  the 8th Workshop on General Purpose Processing Using GPUs}}
  \emph{(\bibinfo{series}{GPGPU-8})}.
\newblock


\bibitem[\protect\citeauthoryear{Rawat, Hong, Ravishankar, Grover, Pouchet,
  Rountev, and Sadayappan}{Rawat et~al\mbox{.}}{2016}]%
        {pact16:rawat}
\bibfield{author}{\bibinfo{person}{Prashant~Singh Rawat},
  \bibinfo{person}{Changwan Hong}, \bibinfo{person}{Mahesh Ravishankar},
  \bibinfo{person}{Vinod Grover}, \bibinfo{person}{Louis-Noel Pouchet},
  \bibinfo{person}{Atanas Rountev}, {and} \bibinfo{person}{P. Sadayappan}.}
  \bibinfo{year}{2016}\natexlab{}.
\newblock \showarticletitle{Resource Conscious Reuse-Driven Tiling for GPUs}.
  In \bibinfo{booktitle}{\emph{Proceedings of the 2016 International Conference
  on Parallel Architectures and Compilation}} \emph{(\bibinfo{series}{PACT
  '16})}.
\newblock


\bibitem[\protect\citeauthoryear{{Rawat}, {Vaidya}, {Sukumaran-Rajam},
  {Rountev}, {Pouchet}, and {Sadayappan}}{{Rawat} et~al\mbox{.}}{2019}]%
        {ipdps19:artemis}
\bibfield{author}{\bibinfo{person}{P.~S. {Rawat}}, \bibinfo{person}{M.
  {Vaidya}}, \bibinfo{person}{A. {Sukumaran-Rajam}}, \bibinfo{person}{A.
  {Rountev}}, \bibinfo{person}{L. {Pouchet}}, {and} \bibinfo{person}{P.
  {Sadayappan}}.} \bibinfo{year}{2019}\natexlab{}.
\newblock \showarticletitle{On Optimizing Complex Stencils on GPUs}. In
  \bibinfo{booktitle}{\emph{2019 IEEE International Parallel and Distributed
  Processing Symposium (IPDPS)}}.
\newblock


\bibitem[\protect\citeauthoryear{Rawat, Vaidya, Sukumaran-Rajam, Rountev,
  Pouchet, and Sadayappan}{Rawat et~al\mbox{.}}{2019}]%
        {ipdps19:rawat}
\bibfield{author}{\bibinfo{person}{Prashant~Singh Rawat},
  \bibinfo{person}{Miheer Vaidya}, \bibinfo{person}{Aravind Sukumaran-Rajam},
  \bibinfo{person}{Atanas Rountev}, \bibinfo{person}{Louis-No{"e}l Pouchet},
  {and} \bibinfo{person}{P Sadayappan}.} \bibinfo{year}{2019}\natexlab{}.
\newblock \showarticletitle{On Optimizing Complex Stencils on GPUs}.
\newblock  (\bibinfo{year}{2019}).
\newblock


\bibitem[\protect\citeauthoryear{Suriana, Adams, and Kamil}{Suriana
  et~al\mbox{.}}{2017}]%
        {halide-reductions}
\bibfield{author}{\bibinfo{person}{Patricia Suriana}, \bibinfo{person}{Andrew
  Adams}, {and} \bibinfo{person}{Shoaib Kamil}.}
  \bibinfo{year}{2017}\natexlab{}.
\newblock \showarticletitle{Parallel Associative Reductions in Halide}. In
  \bibinfo{booktitle}{\emph{Proceedings of the 2017 International Symposium on
  Code Generation and Optimization}} \emph{(\bibinfo{series}{CGO ’17})}.
\newblock


\bibitem[\protect\citeauthoryear{Verdoolaege, Carlos~Juega, Cohen,
  Ignacio~Gomez, Tenllado, and Catthoor}{Verdoolaege et~al\mbox{.}}{2013}]%
        {taco13:ppcg}
\bibfield{author}{\bibinfo{person}{Sven Verdoolaege}, \bibinfo{person}{Juan
  Carlos~Juega}, \bibinfo{person}{Albert Cohen}, \bibinfo{person}{Jose
  Ignacio~Gomez}, \bibinfo{person}{Christian Tenllado}, {and}
  \bibinfo{person}{Francky Catthoor}.} \bibinfo{year}{2013}\natexlab{}.
\newblock \showarticletitle{Polyhedral Parallel Code Generation for CUDA}.
\newblock \bibinfo{journal}{\emph{ACM Trans. Archit. Code Optim.}}
  \bibinfo{volume}{9}, \bibinfo{number}{4} (\bibinfo{date}{Jan.}
  \bibinfo{year}{2013}).
\newblock


\bibitem[\protect\citeauthoryear{{Wang}, {Xie}, and {Cong}}{{Wang}
  et~al\mbox{.}}{2017}]%
        {ipdps17:wang}
\bibfield{author}{\bibinfo{person}{J. {Wang}}, \bibinfo{person}{X. {Xie}},
  {and} \bibinfo{person}{J. {Cong}}.} \bibinfo{year}{2017}\natexlab{}.
\newblock \showarticletitle{Communication Optimization on GPU: A Case Study of
  Sequence Alignment Algorithms}. In \bibinfo{booktitle}{\emph{2017 IEEE
  International Parallel and Distributed Processing Symposium (IPDPS)}}.
\newblock


\bibitem[\protect\citeauthoryear{Wolf and Lam}{Wolf and Lam}{1991}]%
        {pldi91:wolf}
\bibfield{author}{\bibinfo{person}{Michael~E. Wolf} {and}
  \bibinfo{person}{Monica~S. Lam}.} \bibinfo{year}{1991}\natexlab{}.
\newblock \showarticletitle{A Data Locality Optimizing Algorithm}. In
  \bibinfo{booktitle}{\emph{Proceedings of the ACM SIGPLAN 1991 Conference on
  Programming Language Design and Implementation}} \emph{(\bibinfo{series}{PLDI
  '91})}.
\newblock


\bibitem[\protect\citeauthoryear{Wolfe}{Wolfe}{1989}]%
        {wolfe-tiling}
\bibfield{author}{\bibinfo{person}{M. Wolfe}.} \bibinfo{year}{1989}\natexlab{}.
\newblock \showarticletitle{More Iteration Space Tiling}. In
  \bibinfo{booktitle}{\emph{Proceedings of the 1989 ACM/IEEE Conference on
  Supercomputing}} \emph{(\bibinfo{series}{Supercomputing ’89})}.
\newblock


\bibitem[\protect\citeauthoryear{Wolfe}{Wolfe}{1994}]%
        {wolfe-dependence-vector}
\bibfield{author}{\bibinfo{person}{Michael Wolfe}.}
  \bibinfo{year}{1994}\natexlab{}.
\newblock \showarticletitle{The Definition of Dependence Distance}.
\newblock \bibinfo{journal}{\emph{ACM Trans. Program. Lang. Syst.}}
  (\bibinfo{year}{1994}).
\newblock


\bibitem[\protect\citeauthoryear{Zhao and Cohen}{Zhao and Cohen}{2019}]%
        {taco2019:flextended-tiles}
\bibfield{author}{\bibinfo{person}{Jie Zhao} {and} \bibinfo{person}{Albert
  Cohen}.} \bibinfo{year}{2019}\natexlab{}.
\newblock \showarticletitle{Flextended Tiles: A Flexible Extension of
  Overlapped Tiles for Polyhedral Compilation}.
\newblock \bibinfo{journal}{\emph{ACM Trans. Archit. Code Optim.}}
  (\bibinfo{year}{2019}).
\newblock


\bibitem[\protect\citeauthoryear{Zheng, Oh, Zhai, Shen, Yi, and Chen}{Zheng
  et~al\mbox{.}}{2017}]%
        {Versapipe}
\bibfield{author}{\bibinfo{person}{Zhen Zheng}, \bibinfo{person}{Chanyoung Oh},
  \bibinfo{person}{Jidong Zhai}, \bibinfo{person}{Xipeng Shen},
  \bibinfo{person}{Youngmin Yi}, {and} \bibinfo{person}{Wenguang Chen}.}
  \bibinfo{year}{2017}\natexlab{}.
\newblock \showarticletitle{Versapipe: A Versatile Programming Framework for
  Pipelined Computing on GPU}. In \bibinfo{booktitle}{\emph{Proceedings of the
  50th Annual IEEE/ACM International Symposium on Microarchitecture}}
  \emph{(\bibinfo{series}{MICRO-50 ’17})}.
\newblock


\bibitem[\protect\citeauthoryear{Zheng, Oh, Zhai, Shen, Yi, and Chen}{Zheng
  et~al\mbox{.}}{2019}]%
        {HiWayLib}
\bibfield{author}{\bibinfo{person}{Zhen Zheng}, \bibinfo{person}{Chanyoung Oh},
  \bibinfo{person}{Jidong Zhai}, \bibinfo{person}{Xipeng Shen},
  \bibinfo{person}{Youngmin Yi}, {and} \bibinfo{person}{Wenguang Chen}.}
  \bibinfo{year}{2019}\natexlab{}.
\newblock \showarticletitle{HiWayLib: A Software Framework for Enabling High
  Performance Communications for Heterogeneous Pipeline Computations}. In
  \bibinfo{booktitle}{\emph{Proceedings of the Twenty-Fourth International
  Conference on Architectural Support for Programming Languages and Operating
  Systems}} \emph{(\bibinfo{series}{ASPLOS ’19})}.
\newblock


\end{thebibliography}

\iftrue
\newpage
\appendix
\section{Generalized Hybrid Tiling Algorithm}
\label{appendix}

In this appendix, we extend the Hybrid Tiling algorithm of Section~\ref{sec:hybrid-tiling} to support hybrid tiling for groups with 3-D computations and any dimension as split dimension. 
Our algorithm uses dependence vectors to determine which register to load from which thread of warp. In this section, we follow following notations. We refer to the right hyperplanes of given group as $\phi_{ri}$ in $i^{th}$ dimension and the dependence vector between a producer-consumer pair in $i^{th}$ dimension as $\phi_i$. (PolyMage's overlapped tiling algorithm ensures that for a group, we should always have $\phi_i \leq \phi_{ri}$.) Let $i$ is the split dimension, register tile size in $i^{th}$ dimension is $R_i$, a thread's lane id in the $i^{th}$ dimension is $t_i$, and $r_i$ is the current parallelogram tile the thread is executing. 

The type of load a thread perform (shown in Figure~\ref{fig:cuda-blur-reg-tile}) depends on the value of $t_i$, $|\phi_i - \phi_{ri}|$, and $W_i$. Since each parallelogram tile is of size $W_i$, we need to divide the linear register address relative to $t_i$, which is $|\phi_i - \phi_{ri}|$, into a two dimensional relative addresses that contains: (i) lane id of source thread, and (ii) the index in register array. This division is performed on basis of three cases. (1) If the thread needs to read a register from current parallelogram tile, then the thread will read register, $r_i - 1 - \frac{|\phi_i - \phi_{ri}|}{W_i}$ from a thread with lane id, $t_i - (|\phi_i - \phi_{ri}| \bmod W_i)$. (2) If the thread needs to read from previous parallelogram tiles, then the thread will read register $r_i - 1 - \frac{|\phi_i - \phi_{ri}|}{W_i}$ from a source thread with lane id $W_i - 1 - (|\phi_i - \phi_{ri}| \bmod W_i)$. (3) Otherwise the thread will read from shared memory.
Hence, first $|\phi_i - \phi_{ri}| \bmod W_i$ threads of a warp either loads from shared memory (Type~\textcircled{2}) if register $r_i - \frac{|\phi_i - \phi_{ri}|}{W_i}$ does not exists, i.e., it is less than zero, or from another thread's register in previous parallelogram tile (Type~\textcircled{4}), while the remaining threads of the warp reads from register of thread of current parallelogram tile (Type~\textcircled{3}) or from register of current thread (Type~\textcircled{1}).

Algorithm~\ref{algo:full-hybrid-tiling} is our general hybrid tiling algorithm that generates code based on the above idea. \textsc{3-D-HybridTiling} function takes four arguments: (i) the group G, (ii) size of register tile as the fraction of warp overlapped tile, (iii) tile sizes in each dimension, and (iv) warp sizes in each dimension. This function generates the code for hybrid tiling by replacing each producer access with one of CUDA code given in Figure~\ref{fig:full-ht-code-cases}. We first determine a split dimension that must have a tile size greater than 1 (lines~\ref{line:algo:hybrid-tiling:split-dim-start}--\ref{line:algo:hybrid-tiling:split-dim-end}). We then create parallelogram tiles in the split dimension with each tile of size equal to the warp size (line~\ref{line:algo:hybrid-tiling:parallelogram-tiles}). We determine the size of register tiles in each dimension (line~\ref{line:algo:hybrid-tiling:reg-tile-size-start}--\ref{line:algo:hybrid-tiling:reg-tile-size-end}). For each stage in the group, we first generate shared memory tiles using existing PolyMage's compiler and then generate register tiles using \textsc{GenRegTile} function.
 
\textsc{GenRegTile} function takes six arguments: (i) a stage of the group, (ii) the split dimension, (iii) right hyperplanes of the group in each dimension, (iv) register tile sizes in each dimension, and (v) warp sizes in each dimension. We unroll the register tile loop and then replace each consumer store with a register store, and replace each producer load with code in one of the two cases in Figure~\ref{fig:full-ht-code-cases} (lines\ref{algo:full-hybrid-tiling:unroll-loops}--\ref{line:algo:full-hybrid-tiling:unroll-loops-end}). We now explain each of the two cases in detail. (1) If $i$ is the split dimension and $\phi_i - \phi_{ri} \bmod W_i$ then code in Figure~\ref{fig:full-ht-case-1-code} is generated, i.e., the producer load was stored by the current thread in its register (Type~\textcircled{1}) at index $R_i - \frac{|\phi_i - \phi_{ri}|}{W_i}$. (2) Otherwise, we generate code in Figure~\ref{fig:full-ht-case-2-code}.
In the code, there are two type of source lane ids: (i) \texttt{currTileSrcLane} represents the source lane id of thread in current parallelogram tile and (ii) \texttt{prevTileSrcLane} represents the source lane id of thread in previous parallelogram tile. Both source lane ids are linearized by multiplying them with warp sizes. We first generate a Type~\textcircled{3} load that is performed by all threads. Then a conditional is generated that decides if first $|\phi_i - \phi_{ri}| \bmod W_i$ threads of warp should perform Type~\textcircled{2} or Type~\textcircled{1} load.

\begin{figure}[h]
  \footnotesize
\begin{subfigure}[t]{\linewidth}
    \begin{lstlisting}[language=C,escapechar=|]
val = Reg_P[z][x-diffPhi.x/warpSize.x][c*y + d]
  \end{lstlisting}
  \caption{Code for a register access from same thread is generated when the source lane ID is the current lane ID, i.e., $\phi_x - \phi_{rx}| \bmod W_x$ is zero (Type \textcircled{1}). \label{fig:full-ht-case-1-code}}
\end{subfigure}

\begin{subfigure}[t]{\linewidth}
  \begin{lstlisting}[language=C,escapechar=|]
diffModWarp.x = diffPhi.x%warpSize.x
diffModWarp.y = diffPhi.y%warpSize.y
diffModWarp.z = diffPhi.z%warpSize.z
prevTileSrcLane = (warpSize.x - 1 - diffModWarp.x)+
  (warpSize.y - 1 - diffModWarp.y)*warpSize.x+
  (warpSize.z - 1 - diffModWarp.z)*warpSize.y*warpSize.x;
currTileSrcLane = (laneId.x - diffModWarp.x)+
  (laneId.y - diffModWarp.y)*warpSize.x +
  (laneId.z - diffModWarp.z)*warpSize.y*warpSize.x;
/*Type 3:*/ 
  val = __shfl_sync(getMask(), 
          Reg_P[z][x-diffPhi.x/warpSize.x][c*y+d], 
          currTileSrcLane);
if (x - 1 - diffModWarp.x < 0)
  /*Type 2:*/ val = Sh_P[z][a*x+b][c*y+d];
else if (laneId.x - diffModWarp.x < 0)
  /*Type 4:*/ 
  val = __shfl_sync(getMask(), 
    Reg_P[z][x-1-diffPhi.x/warpSize.x][c*y+d],
    prevTileSrcLane);

\end{lstlisting}
\caption{Above code is generated when $x$ is split dimension and $\phi_x-\phi_{rx}\neq 0$. When the sum of lane index and $|\phi_x-\phi_{rx}|$ is less than zero, then value is accessed from register of last $|\phi_x - \phi_{rx}|$ threads of previous parallelogram tile (Type \textcircled{4}) otherwise value is accessed from register of thread in the same parallelogram tile (Type \textcircled{3}).  \label{fig:full-ht-case-2-code}}
\end{subfigure}

\caption{\small Two code generation cases for a producer \texttt{p[a*x+b][c*y+d]} that generates all four load types of Figure~\ref{fig:regtiling} when $x$ is split dimension. Code generated for any other split dimension can be obtained by replacing $x$ with the other dimension in the \texttt{if} statements and updating register array accesses for the other dimension. Code generator replaces \texttt{p[a*x+b][c*y+d]} with \texttt{val}. \texttt{Reg\_P} is the register array storing register tile. \texttt{laneId.x} and \texttt{laneId.y} are the lane indices in x and y dimensions of the current thread. \texttt{warpSize.x}, \texttt{warpSize.y}, and \texttt{warpSize.z} are the warp sizes in x, y, and z dimensions. \texttt{diffPhi.x} is the value of $|\phi_x - \phi_{rx}|$. \texttt{diffPhi.y} is the value of $|\phi_y - \phi_{ry}|$. 
\texttt{diffPhi.z} is the value of $|\phi_z - \phi_{rz}|$. 
\label{fig:full-ht-code-cases}
}
\end{figure}

\begin{algorithm}[H]
  \small
  \caption{Generalized Hybrid Tiling}\label{algo:full-hybrid-tiling}
  \begin{algorithmic}[1]
      \Function{GenRegTile} {H, $\phi_{rx}$, $\phi_{ry}$,  $\phi_{rz}$, R$_x \times$ R$_y \times$ R$_z$, W$_x \times $W$_y \times$ W$_z$}
          \ForAll {\{x, y, z\} $\in$ [1, \ldots R$_x$] $\times$ [1, \ldots R$_y$] $\times$ [1, \ldots R$_z$]} \label{algo:full-hybrid-tiling:unroll-loops}
            \State{Let iteration \{x, y, z\} be}
            \State{\texttt{H[z][x][y] = f(P[z][a*x+b][c*y+d], ...)}}
            \State{Store \texttt{H[z][x][y]} in register \texttt{Reg\_H[z][x][y]}} \label{line:algo:hybrid-tiling:store-to-var}
            \ForAll{loads \texttt{p[z][a*x+b][c*y+d]} $\in f$} \label{line:algo:hybrid-tiling:update-producer-load-start}
              \State{$\phi_x,\phi_y,\phi_z$ = dependence vectors between \texttt{P[z][a*x+b][c*y+d]} and \texttt{H[z][x][y]}}\label{line:algo:hybrid-tiling:phi_x}
              \If{$\phi_i$ == $\phi_{ri}$}
                \State{Generate Type\textcircled{1} code in Figure~\ref{fig:case-1-code}} \label{line:algo:hybrid-tiling:case-1}
              \Else 
                \State{Generate Type \textcircled{3} and \textcircled{4} code from Figure~\ref{fig:case-3-code}}
              \EndIf
            \EndFor\label{line:algo:full-hybrid-tiling:unroll-loops-end}
          \EndFor
      \EndFunction
      \Function{3-D-HybridTiling} {G, fracReg, T$_x \times$ T$_y \times$ T$_z$, W$_x \times$ W$_y \times $W$_z$}
        \State{splitDim = \textit{a dimension with tile size greater than 1}} \label{line:algo:hybrid-tiling:split-dim-start}
        \State{\textit{If no split dimension exists then} \textbf{return}}
        \State{\textit{Let $\phi_{ri}$ be the right hyperplanes of G in $i$ dimension, where $i \in \{x, y, z\}$}}
        \State{\textit{Let} splitDim is the $i \in \{x, y, z\}$.} \label{line:algo:hybrid-tiling:split-dim-end}
        \State{\textit{Create parallelogram tiles in $i$-dim of size W$_i$ parallel to $\phi_{ri}$}} \label{line:algo:hybrid-tiling:parallelogram-tiles}
        \State{R$_i$ $\gets$ T$_i \times $ fracReg, S$_i$ $\gets$ T$_i \times $ (1 - fracReg)} \label{line:algo:hybrid-tiling:reg-tile-size-start}
        \ForAll {$j \in \{x, y, z\}$ s.t. $j \neq i$}
          \State{R$_j$  $\gets$ S$_j$ $\gets$ T$_j$}
        \EndFor\label{line:algo:hybrid-tiling:reg-tile-size-end}
        \ForAll {H $\in$ G}
          \State{\textit{Gen. Shared Mem Tile with tile size S$_x \times $S$_y \times $S$_z$}} \label{line:algo:hybrid-tiling:gen-shared-mem-tile}
          \State {\textsc{GenRegTile}(H, $\phi_{rx}$, $\phi_{ry}$, $\phi_{rz}$, R$_x \times$ R$_y \times$ R$_z$, W$_x \times $W$_y \times$ W$_z$)} \label{line:algo:hybrid-tiling:gen-reg-tile}
        \EndFor
      \EndFunction
  \end{algorithmic}
\end{algorithm}
\fi
\end{document}